\documentclass[twocolumn]{aastex61_2}

\usepackage{amssymb,multirow,morefloats,amsmath}
\usepackage[figuresright]{rotating}
\usepackage{threeparttable}
\usepackage{longtable}
\usepackage{epstopdf}
\usepackage{graphicx}

\newcommand\T{\rule{0pt}{2.6ex}}       
\newcommand\B{\rule[-1.2ex]{0pt}{0pt}} 
\defcitealias{lehmer10}{L10}
\defcitealias{lehmer16}{L16}
\defcitealias{aird17a}{A17}

\newcommand\aastex{AAS\TeX}

\received{June 1, 2018}
\revised{July 26, 2018}
\accepted{August 1, 2018}
\submitjournal{ApJ}

\shorttitle{\aastex\ AGN and XRB populations in COSMOS}
\shortauthors{Fornasini et al.}

\begin{document}

\title{Low-luminosity AGN and X-ray binary populations in COSMOS star-forming galaxies}

\correspondingauthor{Francesca M. Fornasini}
\email{francesca.fornasini@cfa.harvard.edu}

\author{Francesca M. Fornasini}
\affil{Harvard-Smithsonian Center for Astrophysics, 60 Garden Street, Cambridge, MA 02138, USA}

\author{Francesca Civano}
\affiliation{Harvard-Smithsonian Center for Astrophysics, 60 Garden Street, Cambridge, MA 02138, USA}

\author{Giuseppina Fabbiano}
\affiliation{Harvard-Smithsonian Center for Astrophysics, 60 Garden Street, Cambridge, MA 02138, USA}

\author{Martin Elvis}
\affiliation{Harvard-Smithsonian Center for Astrophysics, 60 Garden Street, Cambridge, MA 02138, USA}

\author{Stefano Marchesi}
\affiliation{Department of Physics \& Astronomy, Clemson University, Clemson, SC 29634, USA}

\author{Takamitsu Miyaji}
\affiliation{Instituto de Astronom\'ia sede Ensenada, Km. 103 Carret. Tijuana-Ensenada, Ensenada, 22860 Mexico}
\affiliation{IA-UNAM,PO BOX 439027, San Ysidro, CA 92143-9027, USA (international mailing address)} 

\author{Andreas Zezas}
\affiliation{Foundation for Research and Technology-Hellas, 100 Nikolaou Plastira Street, 71110 Heraklion, Crete, Greece}
\affiliation{Physics Department \& Institute of Theoretical \& Computational Physics, P.O. Box 2208, 71003 Heraklion, Crete, Greece}
\affiliation{Harvard-Smithsonian Center for Astrophysics, 60 Garden Street, Cambridge, MA 02138, USA}



\begin{abstract}

We present an X-ray stacking analysis of $\sim$75,000 star-forming galaxies between $0.1<z<5.0$ using the \textit{Chandra} COSMOS Legacy survey to study the X-ray emission of low-luminosity active galactic nuclei (AGN) and its connection to host galaxy properties.  The stacks at $z<0.9$ have luminosity limits as low as $10^{40}-10^{41}$ erg s$^{-1}$, a regime in which X-ray binaries (XRBs) can dominate the X-ray emission.  Comparing the measured luminosities to established XRB scaling relations, we find that the redshift evolution of the luminosity per star formation rate (SFR) of XRBs depends sensitively on the assumed obscuration and may be weaker than previously found.  The XRB scaling relation based on stacks from the \textit{Chandra} Deep Field South overestimates the XRB contribution to the COSMOS high specific SFR (sSFR) stacks, possibly due to a bias affecting the CDF-S stacks because of their small galaxy samples.  After subtracting the estimated XRB contribution from the stacks, we find that most stacks at $z>1.3$ exhibit a significant X-ray excess indicating nuclear emission.  The AGN emission is strongly correlated with stellar mass but does not exhibit an additional correlation with SFR.  The hardness ratios of the high-redshift stacks indicate that the AGN are substantially obscured ($N_{\mathrm{H}}\sim10^{23}$ cm$^{-2}$).  These obscured AGN are not identified by IRAC color selection and have $L_X\sim10^{41}-10^{43}$ erg s$^{-1}$, consistent with accretion at an Eddington rate of $\sim10^{-3}$ onto $10^7-10^8 M_{\odot}$ black holes.  Combining our results with other X-ray studies suggests that AGN obscuration depends on stellar mass and an additional variable, possibly the Eddington rate.

\end{abstract}

\keywords{galaxies:active --- galaxies:starburst --- X-rays: galaxies --- X-rays: binaries}



\section{Introduction} 
\label{sec:intro}


A key ingredient of galaxy evolution that is still not fully understood is the relationship between the formation of stars and the growth of the supermassive black hole (BH).  The processes regulating galaxy and BH growth are thought to be linked across cosmic time because of the observed correlations at $z=0$ between BH mass and the large scale properties of galaxies such as stellar mass ($M_*$; e.g., \citealt{magorrian98};
\citealt{ferrarese00}; \citealt{gebhardt00}; \citealt{haring04}; \citealt{mcconnell13}; \citealt{kormendy13}), and the striking resemblance between the cosmic histories of star formation and BH accretion (e.g., \citealt{hopkinsbeacom06}; \citealt{silverman08}; \citealt{aird10}). Different physical mechanisms that could trigger BH growth have been proposed, including major galaxy mergers (\citealt{sanders88}; \citealt{dimatteo05}; \citealt{hopkinsp06}), and secular processes driving gas inflow (\citealt{englmaier04}; \citealt{hopkinshernquist06}; \citealt{hopkins10}).  However, uncertainties remain with regards to what extent these different processes contribute to BH growth, and how their relative importance varies with redshift and different levels of BH accretion (see reviews by \citealt{alexander12} and \citealt{heckman14}).  \par
Numerous studies have investigated the relationship between the accretion of active galactic nuclei (AGN) and the star formation rates (SFRs) of their host galaxies.  In high-luminosity AGN ($L_{\mathrm{bol}}\gtrsim10^{45}$ erg s$^{-1}$), a strong correlation between the AGN luminosity, a proxy for the BH accretion rate (BHAR), and SFR is observed (e.g. \citealt{lutz08}; \citealt{bonfield11}; \citealt{mor12}; \citealt{rosario12}); major mergers may drive the high SFRs and BHARs in these galaxies.  \par
However, the BHARs of lower luminosity AGN and the SFRs of their host galaxies exhibit at most a weak correlation when compared on a source by source basis (e.g. \citealt{shao10}, \citealt{rosario12}).  Several studies do observe a strong correlation between the SFR and mean BHAR of moderate luminosity AGN binned by SFR (e.g. \citealt{rafferty11}; \citealt{mullaney12}; \citealt{chen13}; \citealt{azadi15}; \citealt{rodighiero15}; \citealt{lanzuisi17}), but when the galaxies are binned by BHAR, there is no correlation between BHAR and mean SFR (\citealt{rosario12}; \citealt{lanzuisi17}), and the mean SFRs of moderate luminosity AGN hosts are consistent with those of inactive galaxies (e.g. \citealt{santini12}; \citealt{bongiorno12}; \citealt{mullaney15}; \citealt{suh17}).  It has been suggested that the apparent contradictions in these observed trends result from the shorter variability timescale of the BHAR when driven by secular processes compared to the galaxy-averaged SFR (\citealt{hickox14}; \citealt{volonteri15}).   Whether BH accretion on average is linked to star formation in moderate luminosity AGN remains a matter of debate, as some studies argue that the BHAR in these sources is more strongly connected to the stellar mass of the host galaxy than its SFR \citep{yang17}.  \par
In addition to AGN variability, a factor which can complicate investigations of the relationship between BH and galaxy growth is obscuration.  Obscured AGN may be missed by surveys which probe the rest-frame UV, optical, or near-IR wavelengths, and in X-ray surveys, where they are more easily detected, they can be mistaken for intrinsically lower-luminosity AGN.  If any systematic trends exist between the obscured AGN fraction, host galaxy properties, or BHAR, the measured relationships between BHAR and SFR may be biased.  While some studies find no correlation between AGN obscuration and SFR (\citealt{rosario12}; \citealt{delmoro16}), others observe such a correlation both in the low-luminosity \citep{castro14} and high-luminosity regimes \citep{chen15}.  Conflicting results also exist on the correlation between AGN obscuration and the specific SFR (sSFR$=$SFR/$M_*$) of the host galaxy.  \citet{juneau13} find that the obscured AGN fraction increases with sSFR, while \citet{lanzuisi17} find the opposite relation and argue that the obscured AGN sample used by \citeauthor{juneau13} is contaminated. \par
Improving our understanding of the connection between BH and galaxy growth requires large galaxy and AGN samples so as to be able to account for AGN variability and to elucidate any trends that may exist between AGN obscuration and host galaxy properties.  The 2.2 deg$^2$ \textit{Chandra} COSMOS-Legacy survey \citep{civano16} and associated multi-wavelength coverage of the COSMOS field offer an excellent opportunity to investigate BH-galaxy evolution.  Some of the aforementioned studies were based on X-ray selected AGN samples from the COSMOS-Legacy survey (\citealt{suh17}, \citealt{lanzuisi17}), which contains 4016 X-ray detected sources \citep{civano16}.  \par
Other studies have pushed below the sensitivity threshold of this survey using X-ray stacking techniques in order to probe low-luminosity AGN.  Through X-ray stacking of early-type galaxies (ETGs), \citet{paggi16} find enhanced AGN emission in ETGs with lower stellar masses, and evidence for highly absorbed AGN emission at $z\sim1.2$.  Performing a similar analysis with dwarf galaxies with $M_*<10^{9.5} M_{\odot}$, \citet{mezcua16} discover an X-ray excess above the expected contribution of X-ray binaries (XRBs), which is consistent with emission from intermediate-mass BHs ($M\sim10^5 M_{\odot}$) that are likely obscured at $z>0.8$.  \par
In this paper, we present a complementary stacking study of the X-ray emission of star-forming galaxies in the COSMOS field, focused on the low and moderate luminosity ($L_X\sim10^{40}-10^{43}$ erg s$^{-1}$) AGN population.  Due to the low average X-ray luminosities reached by our stacks, the XRB contribution can be comparable to or even dominant over the AGN emission.  Some studies have shown that the XRB luminosity per SFR and per stellar mass increases with redshift (\citealt{basu13}; \citealt{lehmer16}, hereafter \citetalias{lehmer16}; \citealt{aird17a}, hereafter \citetalias{aird17a}), a trend which is attributed to the formation of more luminous XRBs in lower-metallicity environments (\citealt{dray06}; \citealt{linden10}; \citealt{fragos13}; \citealt{brorby16}).  Thus, in this paper, we also discuss the constraints we can place on the redshift evolution of XRBs. \par
We describe our star-forming galaxy sample in \S\ref{sec:sample}.  Our X-ray stacking analysis and the spectral models we use to calculate rest-frame X-ray luminosities are described in \S\ref{sec:stacking} and \S\ref{sec:specmodel}, respectively.  We estimate the XRB contribution to the stacked X-ray emission in \S\ref{sec:xrb}, and compare our results to previous studies of XRB scaling relations in \S\ref{sec:xrbdiscussion}.  We discuss the relationship between BH activity and host galaxy properties in \S\ref{sec:agnactivity}, and evidence for an obscured AGN population at $z>1.3$ in \S\ref{sec:obscured}.  In \S\ref{sec:conclusion}, we summarize our conclusions and consider how the next generation of X-ray telescopes could improve our understanding of low-luminosity AGN and XRB populations.  Throughout this work, we assume a $\Lambda$CDM cosmology with $\Omega_{\mathrm{m}}$ = 0.3, $\Omega_{\Lambda}$ = 0.7, and $H_0$ = 70 km s$^{-1}$ Mpc$^{-1}$.

\section{Sample Selection}
\label{sec:sample}
We selected our sample of star-forming galaxies from the COSMOS2015 photometric catalog \citep{laigle16}, which improves on previous COSMOS catalogs by using the second UltraVISTA data release \citep{mccracken12}, deeper IR data from the Spitzer Large Area Survey with Hyper-Suprime-Cam (SPLASH) project, and new Y-band data from Subaru/Hyper-Suprime-Cam \citep{miyazaki12}.  \citet{laigle16} identify star-forming galaxies based on their NUV$-r$ vs. $r-J$ colors \citep{williams09}.  This catalog provides photometric redshifts and galaxy properties (i.e. stellar mass, SFR) based on SED-fitting of the available data from $0.2-8$ $\mu$m using LePhare (\citealt{arnouts02}; \citealt{ilbert06}).  To derive photometric redshifts, spiral and elliptical galaxy templates from \citet{polletta07} and young star-forming galaxy templates from \citet{bruzual03} are used with extinction as a free parameter.  Galaxy properties are determined using the method described in \citet{ilbert15}; in short, galaxy SEDs are fit with synthetic spectra from the Stellar Population Synthesis models of \citet{bruzual03}, adopting a \citet{chabrier03} initial mass function, a combination of exponentially declining and delayed star formation histories, solar and half solar metallicity, and two possible attenuation curves.  \par
While SED-derived SFRs exhibit substantial scatter of $0.3-0.5$ dex when compared to other SFR proxies (i.e. UV+IR or far-infrared indicators), the median SFRs determined by SED-fitting are consistent with the medians determined from other indicators for SFR$<50 M_{\odot}$ yr$^{-1}$ (\citealt{wyuts11}; \citealt{kennicutt12}; \citealt{yang17}).  For SFR$>50 M_{\odot}$ yr$^{-1}$, the SED-derived values may underestimate the true values by about 0.5 dex.  However, since in our stacking analysis, we use the mean and median SFR values in wide SFR bins spanning at least 1 dex, and $<11$\% of the galaxies in our sample have SFR$>50 M_{\odot}$ yr$^{-1}$, the SED-derived SFRs have sufficient accuracy for this work.  \par
\begin{figure}
\hspace{-0.5in}
\includegraphics[angle=270,width=0.6\textwidth]{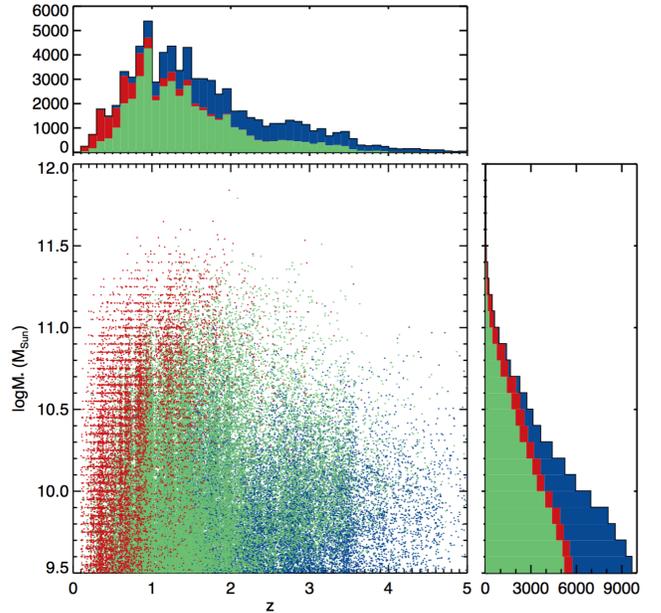}
\caption{Redshift and stellar mass distribution of the COSMOS star-forming galaxy sample used in our stacking analysis.  Galaxies with sSFR$<10^{-9.5}$,  $10^{-9.5}<$sSFR$<10^{-8.5}$, and sSFR$>10^{-8.5}$ yr$^{-1}$ are shown in red, green, and blue, respectively.}
\label{fig:zhist}
\end{figure}
We select galaxies having sSFR$>10^{-11}$ yr$^{-1}$ and stellar mass ($M_*$) greater than $10^{9.5} M_{\odot}$, since dwarf galaxies with lower stellar masses were studied by \citet{mezcua16}.  We exclude any galaxies residing in regions that are flagged by \citet{laigle16} as having saturated pixels or poor photometric quality which is primarily due to contamination by nearby bright stars at optical wavelengths.  \par
We further restrict our sample to galaxies with $i$-band magnitudes $<25$ as a trade-off between maximizing the number of sources (especially at $z>1$) and retaining reasonable precision in the photometric redshifts.  Due to this $i$-magnitude cut, the galaxy sample is $>70\%$ complete for log($M_*/M_{\odot}$)$\geq9.5$ out to $z\approx2$ and $>70\%$ complete for log($M_*/M_{\odot}$)$\geq10.5$ out to $z\approx4$.  We find that excluding the stacks for which the galaxy sample is $<70\%$ complete in stellar mass does not significantly impact our results, probably because there are only 10 such stacks and most are not significant detections and thus do not have much constraining power.  Based on comparisons with spectroscopic redshifts ($z_s$) from multiple spectroscopic surveys in the COSMOS field, \citet{laigle16} find that their photometric redshifts ($z_p$) for star-forming galaxies with $i<25$ have precision $\sigma = \Delta z/(1+z_s) < 0.034$ and a catastrophic failure rate ($\lvert z_p-z_s \rvert/(1+z_s) > 0.15$) of $<10.2$\%.  In our stacking analysis, we select fairly wide ($\Delta z_p \geq 0.2$) redshift bins so that the redshift uncertainties do not significantly impact our results.  Our sample is limited to $0.1<z<5.0$ since only a few dozen galaxies have redshifts outside this range.  Since our goal is to probe the low-luminosity AGN population in star-forming galaxies, we exclude from our sample any galaxies associated with an X-ray detected source in the \textit{Chandra} COSMOS-Legacy survey catalog (\citealt{civano16}; \citealt{marchesi16}).  The sample of sources meeting these criteria consists of 76,845 galaxies.  \par
Finally, as part of our stacking analysis, we impose additional criteria to ensure good X-ray data quality and minimize contamination from bright X-ray sources and diffuse emission (see \S\ref{sec:stacking}). After applying all these selection criteria, our sample consists of 74,904 galaxies.  The redshift and mass distribution of this sample is shown in Figure \ref{fig:zhist}.

\section{Analysis}
\subsection{X-ray Stacking Procedure}
\label{sec:stacking}

With the aim of studying the X-ray emission of galaxies below the COSMOS sensitivity threshold, we perform X-ray stacking analysis making use of the \textit{Chandra} stacking tool CSTACK\footnote{CSTACK (http://lambic.astrosen.unam.mx/cstack/) was developed by Takamitsu Miyaji.} v4.32 \citep{miyaji08}.  For each target, CSTACK provides the net (background-subtracted) count rate in the soft ($0.5-2$ keV) and hard ($2-8$ keV) bands using all 117 observations from the \textit{Chandra} COSMOS-Legacy survey and associated exposure maps.  Since the survey is a highly overlapping mosaic, the same position is observed by multiple observations at different off-axis angles.  Due to the variation of the \textit{Chandra} PSF with off-axis angle, for each observation of an object CSTACK defines a circular source extraction region with size determined by the 90\% encircled counts fraction (ECF) radius ($r_{90}$) (with a minimum of 1$^{\prime\prime}$), thus optimizing the signal-to-noise ratio of the stacked signals.  The background region for each source consists of a $30\times30$ arcsec$^2$ area centered on the object, excluding a 7$^{\prime\prime}$-radius circle around the object and circles around detected X-ray sources with radii dependent on the net X-ray source counts.  For a given object, CSTACK by default only uses observations in which the source is located within 8$^{\prime}$ of the aim point, where $r_{90}<7^{\prime\prime}$.   
\par
\begin{figure}
\centering
\includegraphics[width=0.47\textwidth]{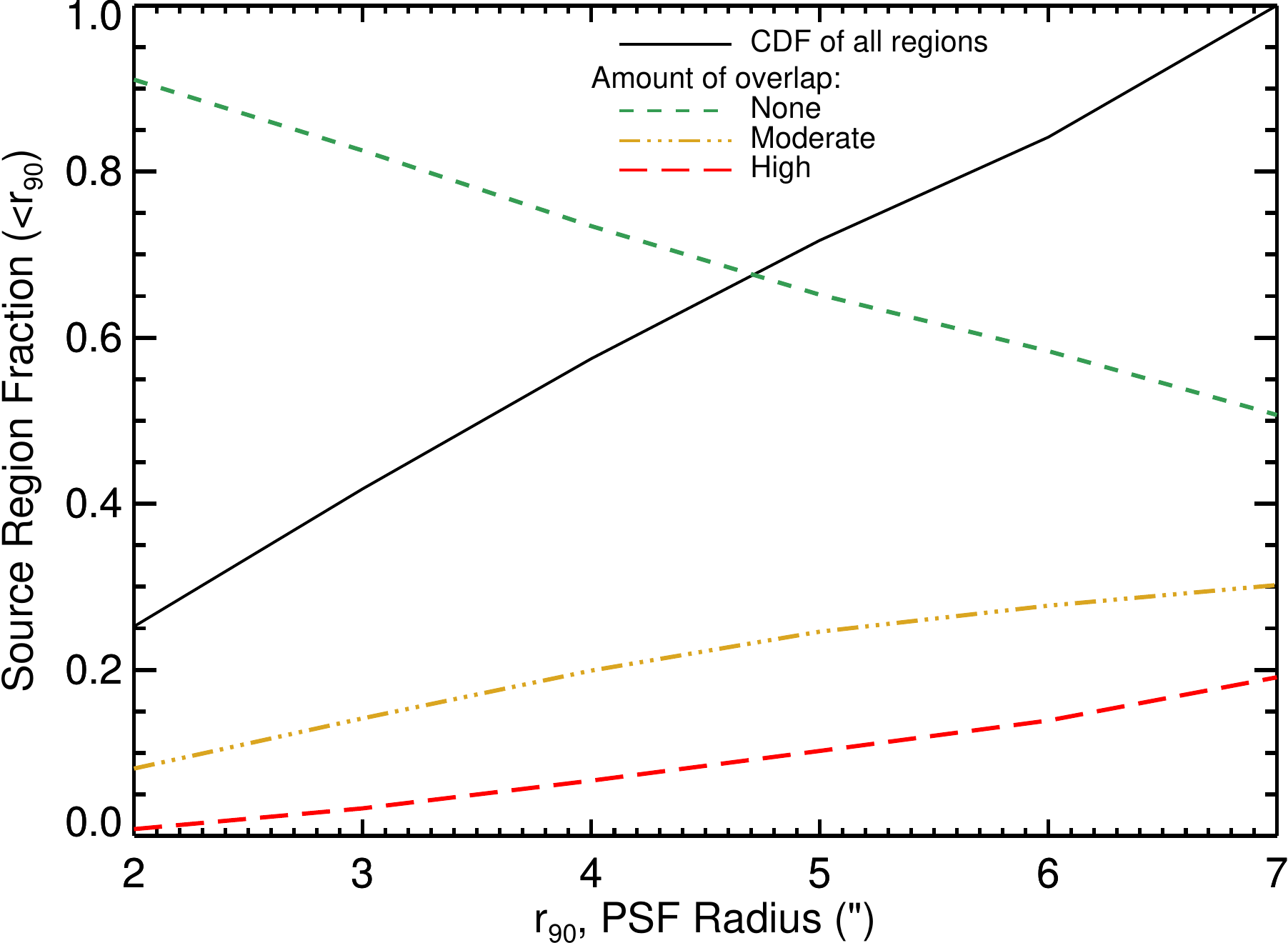}
\caption{The cumulative distribution of source extraction regions for our galaxy sample is shown in black as a function of the 90\% ECF radius.  The green dashed line shows the fraction of source regions with $r_{90}$ smaller than a given value that do not overlap with any other source regions in the parent sample.  The red dashed line shows the fraction of regions for which there is a neighboring source at a distance smaller than $r_{90}$, resulting in a high amount of overlap with another source region.  The gold dash-dotted line shows the fraction of regions for which there is a neighboring source at a distance of $1-2$ times $r_{90}$, resulting in a moderate amount of overlap with another source region.}
\label{fig:contamination}
\end{figure}

 
However, due to the high spatial density of our galaxy sample ($\approx10$ sources per arcmin$^{-2}$), for each object we only make use of observations where $r_{90}<5^{\prime\prime}$ in order to limit bias due to ``double-counting'' the contribution of sources with partially overlapping PSFs.  Figure \ref{fig:contamination} shows the fractions of source extraction regions with $r_{90}$ smaller than a given value having different amounts of overlap with neighboring sources; in cases of high (moderate) overlap, there is a neighboring source at a distance $d<r_{90}$ ($r_{90}<d<2r_{90}$).  For $r_{90}>5^{\prime\prime}$, the fraction of source regions with high amounts of overlap is higher than the fraction with no overlap.  As can be seen in Figure \ref{fig:contamination}, by choosing a maximum $r_{90}$ of $<5^{\prime\prime}$, we still make use of 70\% of the source extraction regions, of which 65\% do not overlap with any of the other source regions, 25\% have moderate levels of overlap, and only 10\% have high levels of overlap. \par
By comparing the net count rate distributions of random locations with different amounts of overlap with sources in our sample, we estimate that the net count rates of our stacks may be overestimated by $\lesssim10\%$ as a result of double-counting sources whose PSFs overlap.  This bias is partially balanced by the fact that a typical background region contains two sources from our sample, leading to an average overestimation of the background count rate of 3\%, which may result in an underestimation of the stacked net count rates of 3-15\%.  Thus, we expect the net effect of these biases resulting from the high spatial density of our sample to be smaller than the statistical errors of $\approx25$\% on the stacked net count rates.  \par
Based on the sizes of star-forming galaxies measured by \citet{shibuya15} as a function of $z$ and $M_*$, the source extraction radii used by CSTACK are larger than the effective radii ($r_E$) of the galaxies at $z<0.6$ in $>99.3$\% of cases, and exceed $2r_E$ for $87$\% of the galaxies at these low redshifts.  At higher redshifts, the percentage of galaxies for which the extraction radius does not exceed $2r_E$ is $<5$\%.  Thus, the X-ray photometric information derived for each galaxy stack should be representative of the average X-ray emission of the galaxies as a whole rather than just the nuclear component. \par
CSTACK flags any source whose photometry may be affected by nearby detected sources from the catalog presented in \citet{civano16}, which is then removed from our sample.  COSMOS sources at $z>0.1$ that are individually detected by \textit{Chandra} are most likely AGN-dominated given their X-ray luminosities ($L_X\gtrsim10^{42}$ erg s$^{-1}$), while galaxies falling below the \textit{Chandra} sensitivity limit may be either AGN or XRB dominated.  In addition, we remove sources which may be affected by soft diffuse emission likely associated with galaxy clusters and groups at the positions listed in Table \ref{tab:clusters}.  \par
Finally, if an individual source is detected at $>3\sigma$ based on the CSTACK-extracted source and background counts, we also remove it from our sample even if it was not detected in the \citet{civano16} catalog; these sources tend to be very close to the detection threshold and thus small changes in how the X-ray background is estimated can push them above the threshold.  We exclude these sources because they can otherwise dominate the stacked signal since we adopt a $3\sigma$ detection threshold for the stacks (see \S\ref{sec:stackedvalues}).  The percentage of sources individually detected at $2-3\sigma$ confidence is 2.7\%, only slightly higher than the percentage expected for a noise distribution (2.1\%).  Excluding these $2-3\sigma$ sources from the stacks reduces the mean luminosities by 2$\sigma$, which is not surprising since doing so biases the net count rate distribution towards lower values; however, if we clip the distribution at both ends, removing the $2-3\sigma$ sources and the 2.1\% of sources with the most negative net count rates, then we recover stacked mean luminosities that are statistically consistent with the values obtained when these sources are included in the stacks.  Thus, it does not appear that our stacks are dominated by a small number of relatively bright sources. \par
\begin{table}
\centering
\footnotesize
\caption{Excluded Regions of Soft Diffuse Emission}
\begin{threeparttable}
\begin{tabular}{ccc} \hline \hline
\T R.A. (J2000) & Decl. (J2000) & Radius \\
\B ($^{\circ}$) & ($^{\circ}$) & ($^{\prime\prime}$) \\
\hline
\T 149.9202 & 2.5195 & 80 \\
150.1990 & 1.6637 & 120 \\
150.4259 & 2.4283 & 80 \\
\B 150.5044 & 2.2249 & 50 \\
\hline
\end{tabular}
\end{threeparttable}
\label{tab:clusters}
\end{table}

\begin{figure*}
\centering
\includegraphics[angle=270,width=1.0\textwidth]{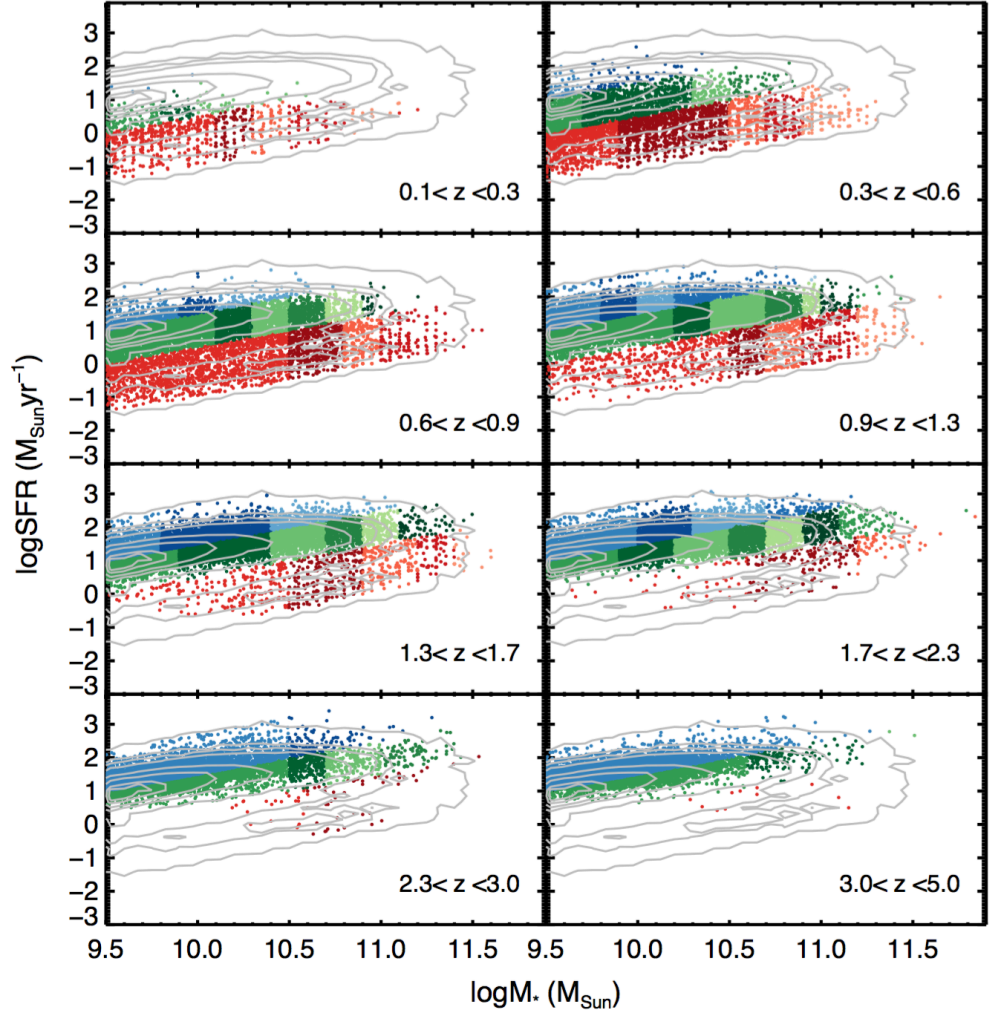}
\caption{COSMOS star-forming galaxy sample plotted by $z$, sSFR, and $M_*$.  Gray contours show the full distribution of SFR and $M_*$ of the 74,904 galaxies in the sample.  Points show individual galaxies split among 92 stacks.  Galaxies with sSFR$<10^{-9.5}$,  $10^{-9.5}<$sSFR$<10^{-8.5}$, and sSFR$>10^{-8.5}$ yr$^{-1}$ are shown in red, green, and blue, respectively; different shades of each color represent different $M_*$ bins.  See Table \ref{tab:properties} for galaxy properties of each stack and Table \ref{tab:photometry} for detection significance and X-ray luminosities of stacks.}
\label{fig:sfms}
\end{figure*}

\subsubsection{Binning of the galaxy sample}
We divide up our galaxy sample into 8 redshift bins, making sure the redshift bins are wider than the photometric redshift uncertainties.  Within each redshift bin, we further split the sample along the star-forming main sequence by both $M_*$ and sSFR in order to study the dependence of X-ray luminosity on galaxy properties. Given that the uncertainties in SED-derived SFRs are substantial (0.3-0.5 dex), we only split the galaxies into three sSFR bins: a low-sSFR bin with log(sSFR/yr$^{-1}$)$<-9.5$, a mid-sSFR bin with $-9.5\leq$log(sSFR)$<-8.5$, and high sSFR-bin with log(sSFR)$>-8.5$.  For reference, the mid-sSFR bin traces the bulk of the star-forming main sequence at $z=1$ (e.g. \citealt{noeske07}), where the redshift distribution of our galaxy sample peaks.  We initially divided the galaxies into mass bins 0.2 dex wide from log($M_*/M_{\odot}$)$=9.5-11.9$ (the 90\% confidence errors on the SED-derived masses are $\leq0.1$ dex), but if a particular stack did not reach our X-ray detection threshold, the mass bin was widened in order to maximize the number of stacked detections.  Ultimately, our sample is divided into 92 stacks, 68 of which exceed our detection threshold (see \S\ref{sec:stackedvalues}).  Figure \ref{fig:sfms} shows the final division of galaxies by $z$, sSFR and $M_*$.  

\begin{figure*}
\centering
\includegraphics[width=1.0\textwidth]{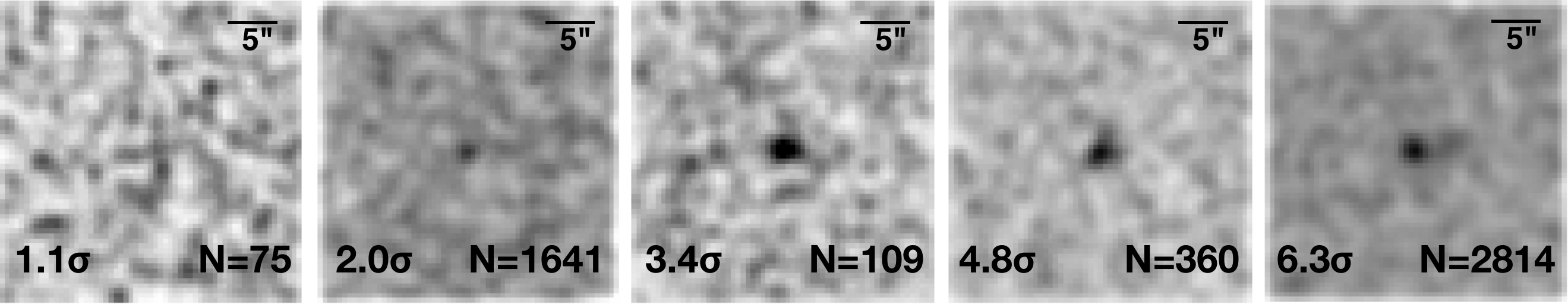}
\caption{Examples of $0.5-2$~keV stacked images of the COSMOS stacks with detection significance shown in the bottom left corner and the number of stacked galaxies shown in the bottom right.  Images have been smoothed with a 3$^{\prime\prime}$-radius Gaussian kernel.}
\label{fig:stackimg}
\end{figure*}

\subsubsection{Calculation of stacked quantities}
\label{sec:stackedvalues}

For each stack, based on the photometric information provided by CSTACK for each galaxy, we calculate the total number of counts within the source regions ($C_{\mathrm{src}}$), the total expected number of background counts within the source regions ($C_{\mathrm{bkg}}$), and the total background-subtracted net source counts ($C_{\mathrm{net}}$).  Then, the probability that the source could be generated by a noise fluctuation of the local background is calculated using the following equation:
\begin{equation}
P(\geq C_{\mathrm{src}}|C_{\mathrm{bkg}};C_{\mathrm{net}}=0) = \sum_{c = C_{\mathrm{src}}}^{\infty}\frac{(C_{\mathrm{bkg}})^c}{c!}e^{-C_{\mathrm{bkg}}} 
\label{eq:probnoise}
\end{equation}
A probability of 0.13\% corresponds to a Gaussian-equivalent 3$\sigma$ detection; we adopt this value as the stacked detection threshold in the soft energy band, which has higher sensitivity than the hard band.  Stacked images of five example stacks are shown in Figure \ref{fig:stackimg} along with their detection significance.  \par
\begin{figure}
\centering
\includegraphics[width=0.47\textwidth]{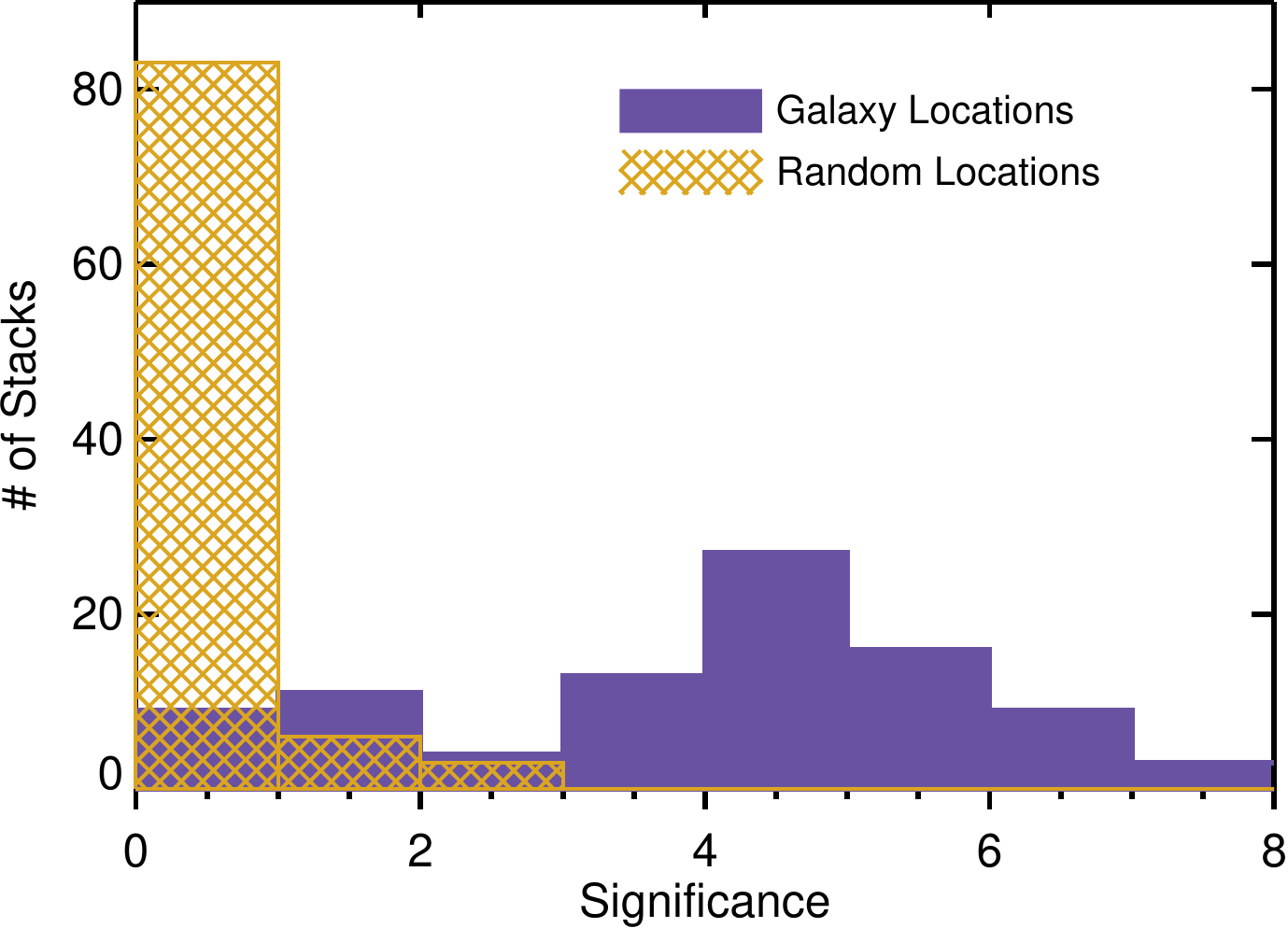}
\caption{Histogram of the detection significance for 92 COSMOS stacks is shown in purple.  The yellow hatched histogram shows the detection significance for the stacks if each galaxy is assigned a random location in the \textit{Chandra} COSMOS field from which its X-ray photometry is extracted.}
\label{fig:sighist}
\end{figure}
For each stack, we compute the exposure-weighted average net count rate, as well as the exposure-weighted means and the standard deviations of the distributions of galaxy properties (i.e. $M_*$, SFR, sSFR, $z$).  The 1$\sigma$ statistical errors on the stacked count rate are calculated using the bootstrapping method; we resample the galaxies in each stack 1,000 times while conserving the number of galaxies and repeat the stacking analysis in order to determine the uncertainties due to sample statistics.  For stacks which are not detected with $>3\sigma$ confidence, we calculate $3\sigma$ upper limits to the net count rate based on the background count rate. The galaxy properties and X-ray photometry of the 92 stacks are provided in Tables \ref{tab:properties} and \ref{tab:photometry}, respectively.  \par
The significance distribution of the stacks is shown by the purple histogram in Figure \ref{fig:sighist}.  We test the robustness of our stacking procedure by assigning to each of the galaxies in each stack a random position within the field of the \textit{Chandra} COSMOS Legacy survey and repeating the stacking analysis. The significance distribution of the stacks based on random positions is shown by the yellow histogram in Figure \ref{fig:sighist}.  The fact that this significance distribution is consistent with what we would expect from random background fluctuations validates our stacking procedure and confirms that our stacked 3$\sigma$ detections are not spurious.  


\subsection{Spectral Dependence of X-ray Luminosities}
\label{sec:specmodel}
\begin{figure*}
\centering
\includegraphics[width=0.85\textwidth]{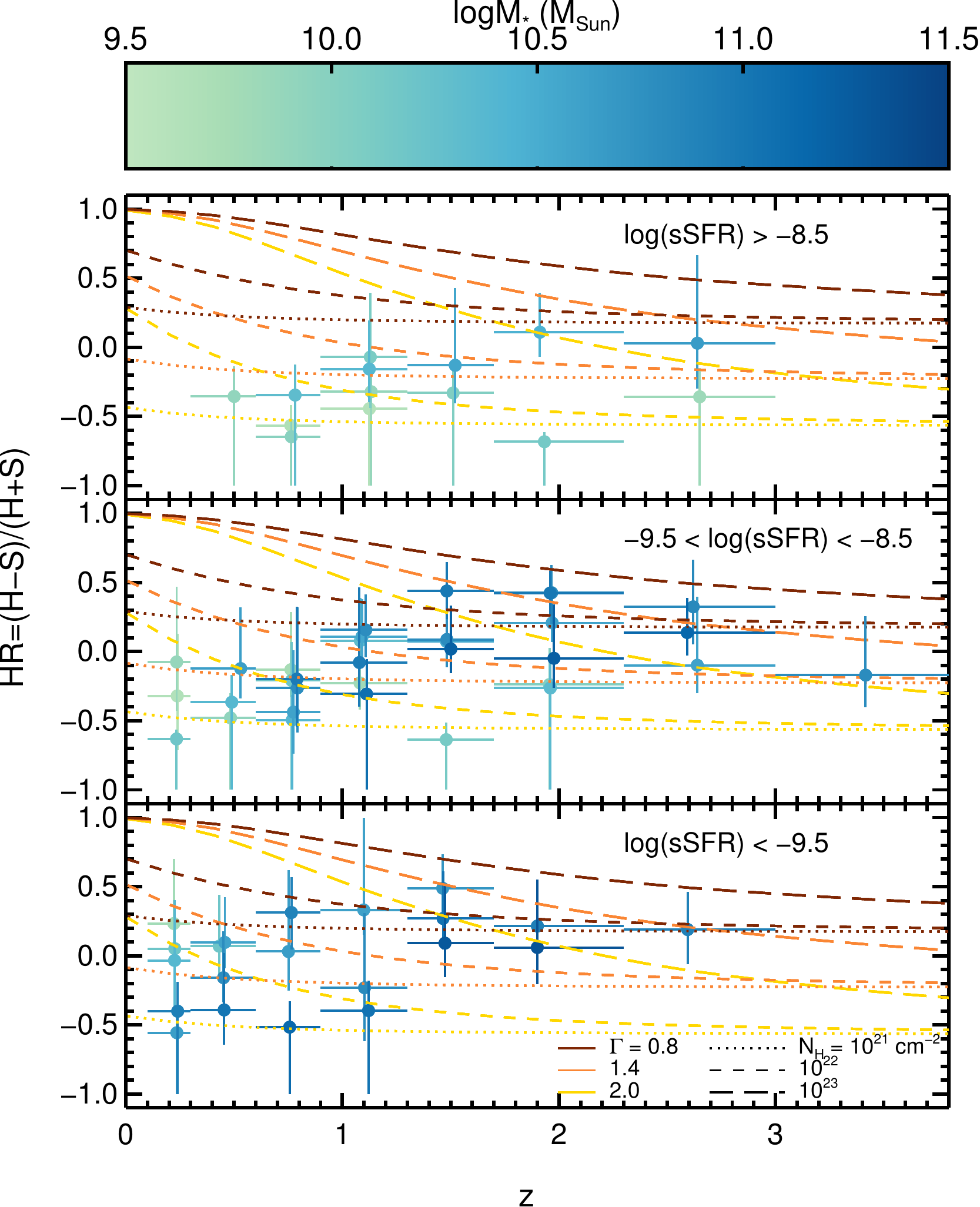}
\caption{Hardness ratios versus redshift for all stacks detected with $\geq3\sigma$ confidence in the $0.5-2$~keV band.  Error bars display 1$\sigma$ uncertainties.  Points are split into panels based on whether the stack has high, mid, or low sSFR and are colored according to the mean stellar mass of the stack.  Colored dotted and dashed lines show the HRs expected for sources with a particular absorbed power-law model; different colors correspond to different photon indices while different line patterns represent different column densities as listed in the legend.  The spectral model used to calculate X-ray luminosities also includes a thermal 1~keV component to represent the hot gas contribution to the X-ray emission.  This component is not accounted for in the model lines shown, since the expected hot gas contribution varies for each stack.  If we removed the hot gas contribution prior to computing the HRs, the HR values of the stacks would increase; the HRs of low-redshift stacks would increase by $\lesssim0.1$, $0.2-0.4$, and $0.4$ dex in the low, mid, and high sSFR regimes, respectively, while the ratios above $z\sim1$ would not be significantly affected.}
\label{fig:hratio}
\end{figure*}

The conversion of stacked count rates into mean X-ray luminosities ($\langle L_X\rangle$) depends on the spectral model assumed.  The X-ray emission of star-forming galaxies originates from three types of sources \citep{fabbiano89}: (i) hot gas, which contributes a diffuse, soft thermal component, (ii) X-ray binaries, and (iii) AGN, the latter two having power-law spectra with $\Gamma\approx1.4-2.0$.  The relative contributions of these three components can vary with galaxy properties (i.e. SFR and $M_*$) and redshift.  While there are not sufficient net counts in each stack to perform spectral fitting, hardness ratios can be calculated for each stack and used to gain insight into an appropriate spectral model to adopt.  For each stack, we calculated the hardness ratio based on the net counts in the $0.5-2$~keV (soft, $S$) and $2-8$~keV (hard, $H$) bands using the Bayesian estimation code BEHR \citep{park06}, which is designed for low count statistics.  The hardness ratio for each stack is defined as $(H-S)/(H+S)$. \par
The hardness ratios for stacks that are detected with $>3\sigma$ significance in the soft band are shown in Figure \ref{fig:hratio} and compared to absorbed power-law spectral models.  As shown in this figure, in the mid and low sSFR stacks, the hardness ratios tend to be higher at $z\gtrsim1$.  The high hardness ratios (HR$\gtrsim0$) at $z>1.3$ suggest that on average the high-redshift X-ray sources either have a flat spectrum ($\Gamma<1$), are highly obscured ($N_{\mathrm{H}}\sim10^{23}$ cm$^{-2}$), or both.  Such flat spectra would be very unusual for luminous XRBs, which are expected to dominate the integrated XRB emission due to the steep luminosity functions of both HMXBs and LMXBs (\citealt{gilfanov04}; \citealt{mineo12a}), but would be consistent with AGN exhibiting a strong Compton reflection component; this strong reflection component is indicative of high obscuration.  Thus, the HRs suggest that at $z>1.3$ our star-forming galaxy stacks with low to mid sSFRs are dominated by obscured AGN.  \citet{paggi16} and \citet{mezcua16} similarly find evidence for obscured AGN at $z\gtrsim1$ by looking at the HR distributions of stacks of early-type galaxies and dwarf galaxies, respectively. \par
The HR errors associated with the high sSFR stacks are so large they make it difficult to determine if there is any redshift evolution.  To increase the signal-to-noise ratio of the high sSFR stacks in the hard band, we adjusted our binning scheme, widening the mass range of each stack.  Even after performing this rebinning, the HR errors remain too large to detect any statistically significant trends with redshift, and the HRs are consistent with the typical spectra of both XRBs and AGN.  \par
In order to select a spectral model appropriate for our stacks, we explore a grid of column density ($N_{\mathrm{H}}=10^{20}-10^{24}$ cm$^{-2}$) and photon index ($\Gamma=0-3$) values for an absorbed power-law model.
In addition to this power-law component representing the XRB and AGN emission, we include a thermal \texttt{apec} model with $kT=1$~keV and Galactic absorption ($N_{\mathrm{H}}=2.6\times10^{20}$ cm$^{-2}$; \citealt{kalberla05}) to represent the hot gas emission; the relative normalization of the thermal and power-law components is set such that the rest-frame $0.5-2$~keV luminosity of the gas component is equal to that predicted by the SFR-$L_{X,\mathrm{gas}}$ correlation based on local star-forming galaxies from \citet{mineo12b}.  The gas contribution to the observed $0.5-2$~keV band is estimated to be $\approx30$\% in the lowest redshift stacks ($z=0.1-0.3$), and it decreases with redshift, down to $<10$\% by $z=0.9$.  If this gas contribution were removed from our stacks prior to calculating the HRs, then the HR values of stacks at $z<1$ would increase by $\lesssim0.1$, $0.2-0.4$, and $0.4$ dex in the low, mid, and high sSFR regimes.  \par
For each combination of $N_{\mathrm{H}}$ and $\Gamma$ in our model parameter space, we use an on-axis \textit{Chandra} ACIS-I auxiliary response file (ARF) from Cycle 14 (which is also used to normalize the CSTACK exposure maps) to calculate the effective area, photon to energy flux conversion factor, and $k$-correction from the observed $0.5-2$~keV or $2-8$~keV band to the rest-frame $2-10$~keV band, a conventional energy band used in many studies.  If the spectral model used is a good representation of the average spectrum for a given stack, the rest-frame $2-10$~keV luminosities derived from the two different observed bands should be consistent.  Therefore, we calculate the difference between the two sets of $2-10$~keV luminosity measurements for all stacks detected with at least $3\sigma$ confidence in the $0.5-2$~keV band and $2\sigma$ confidence in the $2-8$~keV band.  Then, we compute the reduced chi-squared statistic for a model with zero luminosity differences, and determine the 90\% confidence intervals for $N_{\mathrm{H}}$ and $\Gamma$ for which the parameter combination is physically plausible (i.e. a spectrum with $\Gamma\lesssim1$ is also substantially obscured, $N_{\mathrm{H}}>10^{22}$ cm$^{-2}$).  \par
We perform this analysis for the high sSFR stacks separately, and for the combined low-mid sSFR stacks above and below $z=1.3$ independently due to the observed hardness ratio trends.  We find that the high sSFR stacks are well-described by a model with $\Gamma=1.0^{+0.8}_{-0.2}$ and log$N_{\mathrm{H}}=22.0^{+0.8}_{-1.0}$ cm$^{-2}$, while the low-mid sSFR stacks are consistent with $\Gamma=1.2\pm0.5$ and log$N_{\mathrm{H}}=21.8^{+0.6}_{-0.8}$ cm$^{-2}$ at low redshift and with $\Gamma=0.8^{+1.0}_{-0.4}$ and log$N_{\mathrm{H}}=22.4^{+0.8}_{-0.4}$ cm$^{-2}$ at high redshift.   The covariance between $N_{\mathrm{H}}$ and $\Gamma$ is such that the $\chi^2_{\nu}$ statistic for lower values of $\Gamma$ is improved by lower $N_{\mathrm{H}}$ values.  For $\Gamma=1.4$, a value lying within the 90\% confidence range for all the stacks and which is commonly used in AGN studies because it is the photon index of the cosmic X-ray background \citep{deluca04}, the best-fitting $N_{\mathrm{H}}$ values are $10^{22.2}$ cm$^{-2}$ for the high sSFR stacks, $10^{22.0}$ cm$^{-2}$ for the low-$z$, low-mid sSFR stacks, and $10^{23.0}$ cm$^{-2}$ for the high-$z$, low-mid sSFR stacks.  These are the values we adopt for our spectral model.  The derived X-ray luminosities vary by $<0.1$ dex if the best-fitting $\Gamma$ and $N_{\mathrm{H}}$ values are adopted instead or if $\Gamma$ (or $N_{\mathrm{H}}$) is fixed to a different value within the 90\% confidence ranges for all the stacks and the corresponding best-fitting $N_{\mathrm{H}}$ ($\Gamma$) values are used. 
\par 
Throughout this paper, we use absorption-corrected X-ray luminosities that have been corrected for the obscuration in our observationally motivated spectral model.   As a point of comparison, we also calculate X-ray luminosities using a model consisting of a power-law spectrum with $\Gamma=1.9$ subject only to Galactic absorption, which is the model assumed by \citetalias{aird17a}, one of the studies with which we compare our results.  Other studies of XRBs and AGN adopt similar models with little or no host galaxy obscuration that result in luminosity differences of $<0.1-0.2$ dex relative to the \citetalias{aird17a} model (e.g. \citealt{basu13}; \citetalias{lehmer16}; \citealt{mezcua16}).  \par
The X-ray spectra of XRBs and AGN are more complex than a simple absorbed power-law model, including features such as soft disk emission (e.g. \citealt{done07}; \citealt{remillard06}), Compton reflection (e.g. \citealt{george91}), and high-energy cutoffs (e.g. \citealt{gladstone09}; \citealt{fabian15}; \citealt{lubinski16}).  While our data cannot constrain all the parameters associated with these complex physical models, we can use more realistic XRB and AGN spectral templates and determine the effect of spectral models on our results.  For XRBs, we use the average X-ray spectrum of star-forming galaxies observed by \textit{Chandra} and \textit{NuSTAR} from \citetalias{lehmer16} and the spectral models from the XRB population synthesis study of \citet{fragos13}.
For AGN, we use the spectral templates with different levels of obscuration (unobscured, log$N_{\mathrm{H}} = 21.5, 22.5, 23.5, 24.5$) from \citet{gilli07}.
\par
For each of these spectral templates, we compare the rest-frame $2-10$~keV $\langle L_X\rangle$ derived from the observed $0.5-2$ and $2-8$~keV bands, as was done to find the best-fitting parameters for our observationally motivated spectral model.  For both the \citetalias{lehmer16} and \citeauthor{fragos13} templates, we find decent agreement with an average difference of $\langle \Delta L_X\rangle= 0.3$ dex and $\chi_{\nu}^2 = 3.7$ for stacks at $z<0.9$ and the high sSFR stacks.  This agreement is improved if the column density of the model is increased to $N_{\mathrm{H}}\approx10^{22}$ cm$^{-2}$.  All the XRB templates lead to poor agreement between the two sets of derived $\langle L_X\rangle$ values at $z>1.3$ ($\langle\Delta L_X\rangle= 0.8$ dex and $\chi_{\nu}^2 = 22.4$).  The only \citeauthor{gilli07} AGN template that provides decent agreement for all stacks has log$N_{\mathrm{H}} = 22.5$, and yields $\langle\Delta L_X\rangle= 0.2$ dex and $\chi_{\nu}^2 = 3.9$.    This model is in excellent agreement with the stacks at $z<1.3$, resulting in $\langle\Delta L_X\rangle= 0.04$ dex and $\chi_{\nu}^2 = 1.5$.  However, for stacks at $z>1.3$, the AGN template with log$N_{\mathrm{H}} = 23.5$ is more consistent with our stacks ($\langle\Delta L_X\rangle= 0.2$ dex and $\chi_{\nu}^2 = 4.9$ compared to $\langle\Delta L_X\rangle= 0.4$ dex and  $\chi_{\nu}^2 = 6.5$ for log$N_{\mathrm{H}} = 22.5$).  Thus, applying these more complex spectral templates leads to the same conclusions as our simpler absorbed power-law model: at $z\lesssim1$, our stacks are consistent with both XRB and AGN spectra with $N_{\mathrm{H}}\sim10^{22}$ cm$^{-2}$, while at $z\gtrsim1$, our stacks are likely dominated by obscured AGN with $N_{\mathrm{H}}\sim10^{23}$ cm$^{-2}$. \par
Since the AGN template with log$N_{\mathrm{H}} = 22.5$ is in agreement with all our stacks, we have tested whether using this template rather than our observationally motivated spectral model impacts the results described in \S\ref{sec:discussion}.  Apart from increasing the mean X-ray luminosities of stacks at $z<0.9$ by $0.2-0.3$ dex, adopting this AGN template as our spectral model does not significantly change any of our results.  
\par
\begin{figure*}
\centering
\includegraphics[width=0.85\textwidth]{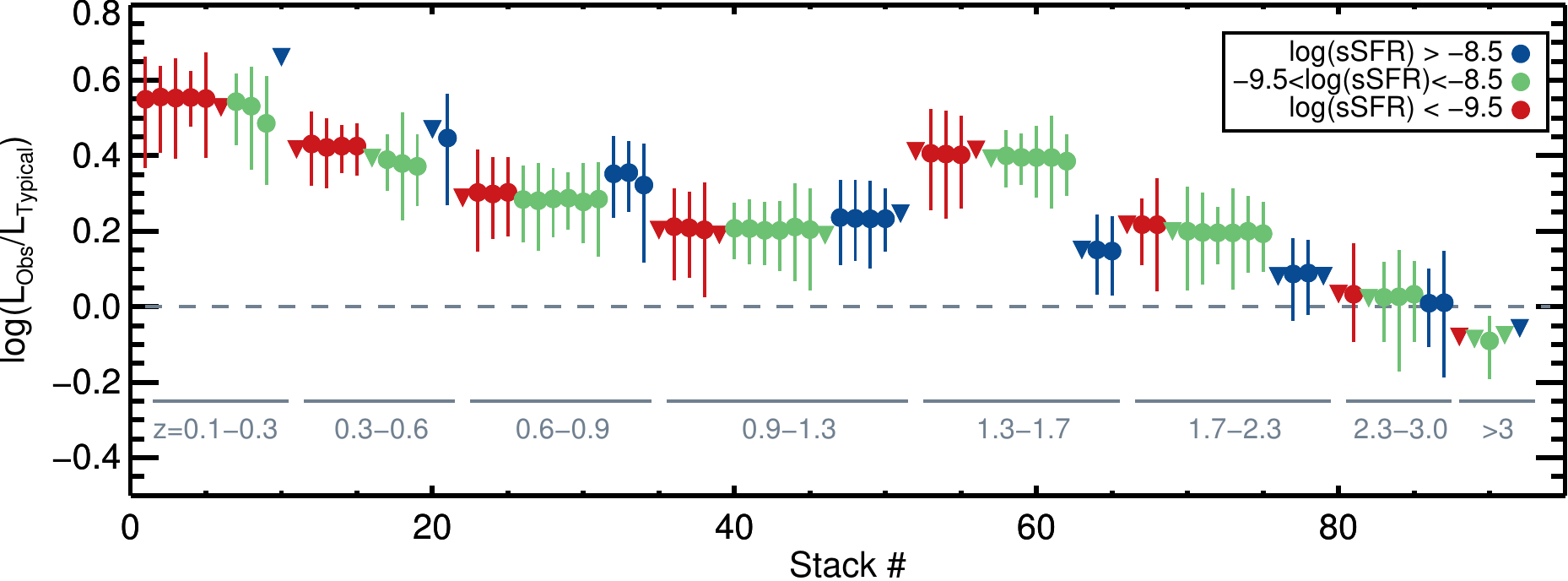}
\caption{Difference between rest-frame $2-10$~keV luminosities calculated for the galaxy stacks using our observationally motivated spectral model based on the hardness ratios and a simpler spectral model with constant $\Gamma=1.9$ and only Galactic absorption.  High, mid, and low sSFR stacks are shown with blue, green, and red symbols, respectively.  Error bars represent 1$\sigma$ statistical uncertainties for stacks detected with $>3\sigma$ confidence; triangle symbols represent stacks with $<3\sigma$ significance.}
\label{fig:lumdiff}
\end{figure*}
As shown in Figure \ref{fig:lumdiff}, the difference between the X-ray luminosities derived using our observationally motivated model and the typical model is $0.1-0.6$ dex.  The fact that this difference is larger than the average statistical luminosity uncertainty of 0.1 dex reveals the importance of using information from multiple X-ray bands to constrain the spectral model when calculating X-ray luminosities.  The difference in the derived $\langle L_X\rangle$ is primarily driven by the higher level of obscuration in the observationally motivated spectral model.  The discrepancy between the derived luminosities almost disappears at $z\sim2-5$ where the $k$-correction is less significant since the observed $0.5-2$ keV band corresponds almost directly to the rest frame $2-10$~keV band. \par
We cannot determine whether the obscuring material is confined to the galaxy nucleus, and thus primarily affects the AGN emission, or if it is spread throughout the galaxy, affecting the XRB and AGN emission to a similar extent. X-ray studies of nearby star-forming galaxies with \textit{Chandra} and \textit{NuSTAR} find that bright XRBs which dominate the X-ray emission above 2~keV are obscured by gas within their host galaxies with $N_{\mathrm{H}}\sim10^{21}-10^{22}$ cm$^{-2}$ (\citealt{wik14}; \citealt{lehmer15}; \citealt{yukita16}).  Thus, the average column densities we measure for our stacks at $z<1.3$ are towards the high end of values measured for XRBs in nearby galaxies but they are not extreme outliers, in particular when one considers that the nearby galaxies that have been studied are mostly face-on while our galaxy sample is unbiased with respect to galaxy inclination.   \par
\begin{figure*}
\centering
\includegraphics[width=0.85\textwidth]{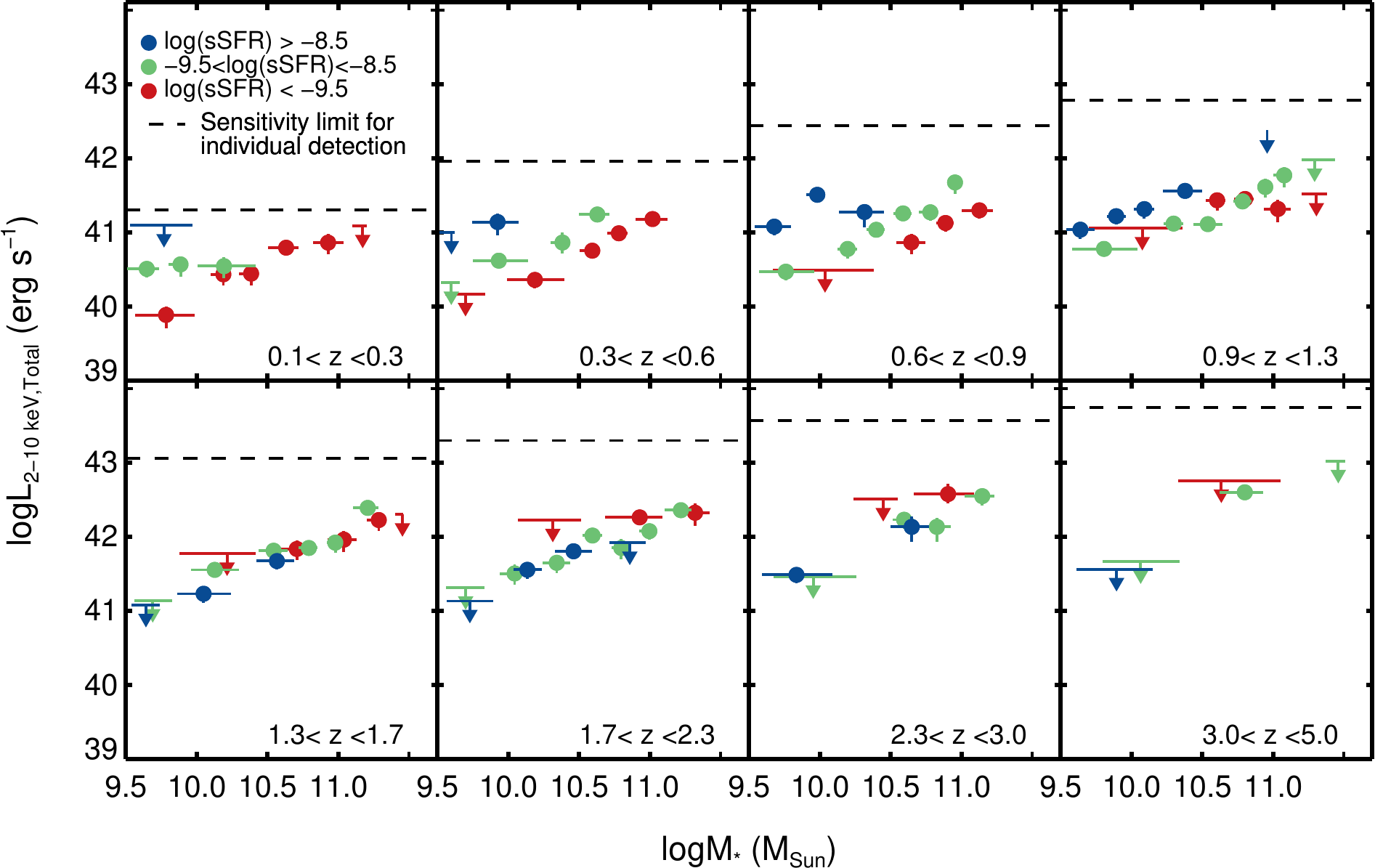}
\caption{Rest-frame $2-10$~keV luminosities of the galaxy stacks versus $M_*$ split into different redshift panels.  $\langle L_X\rangle$ is calculated using our observationally motivated spectral model in which $\Gamma=1.4$ and $N_{\mathrm{H}}$ varies with $z$ and sSFR based on trends observed in the hardness ratios. High, mid, and low sSFR stacks are shown with blue, green, and red symbols, respectively. Stacks detected with $>3\sigma$ confidence are represented by circles with 1$\sigma$ vertical error bars.  The $3\sigma$ upper limits for stacks which do not meet our detection criteria are represented by arrows.  Horizontal error bars represent the standard deviation of the mass distribution of the galaxies in each stack.  The sensitivity limit of the \textit{Chandra} COSMOS Legacy survey is shown by a dashed line in each panel.}
\label{fig:lumtot}
\end{figure*}

The mean unabsorbed rest-frame $2-10$~keV luminosities calculated using our observationally motivated spectral model are shown in Figure \ref{fig:lumtot}.  As shown in this Figure, the stacks allow us to probe luminosities $1-2$ orders of magnitude below the \textit{Chandra} COSMOS survey sensitivity limit.  Some stacks at $z<1.3$ reach luminosity limits as low as $10^{40}-10^{41}$ erg s$^{-1}$, a regime in which XRBs can significantly contribute or even dominate the X-ray emission.

\subsection{Estimating the X-ray binary contribution}
\label{sec:xrb}

\begin{figure*}
\centering
\includegraphics[width=0.85\textwidth]{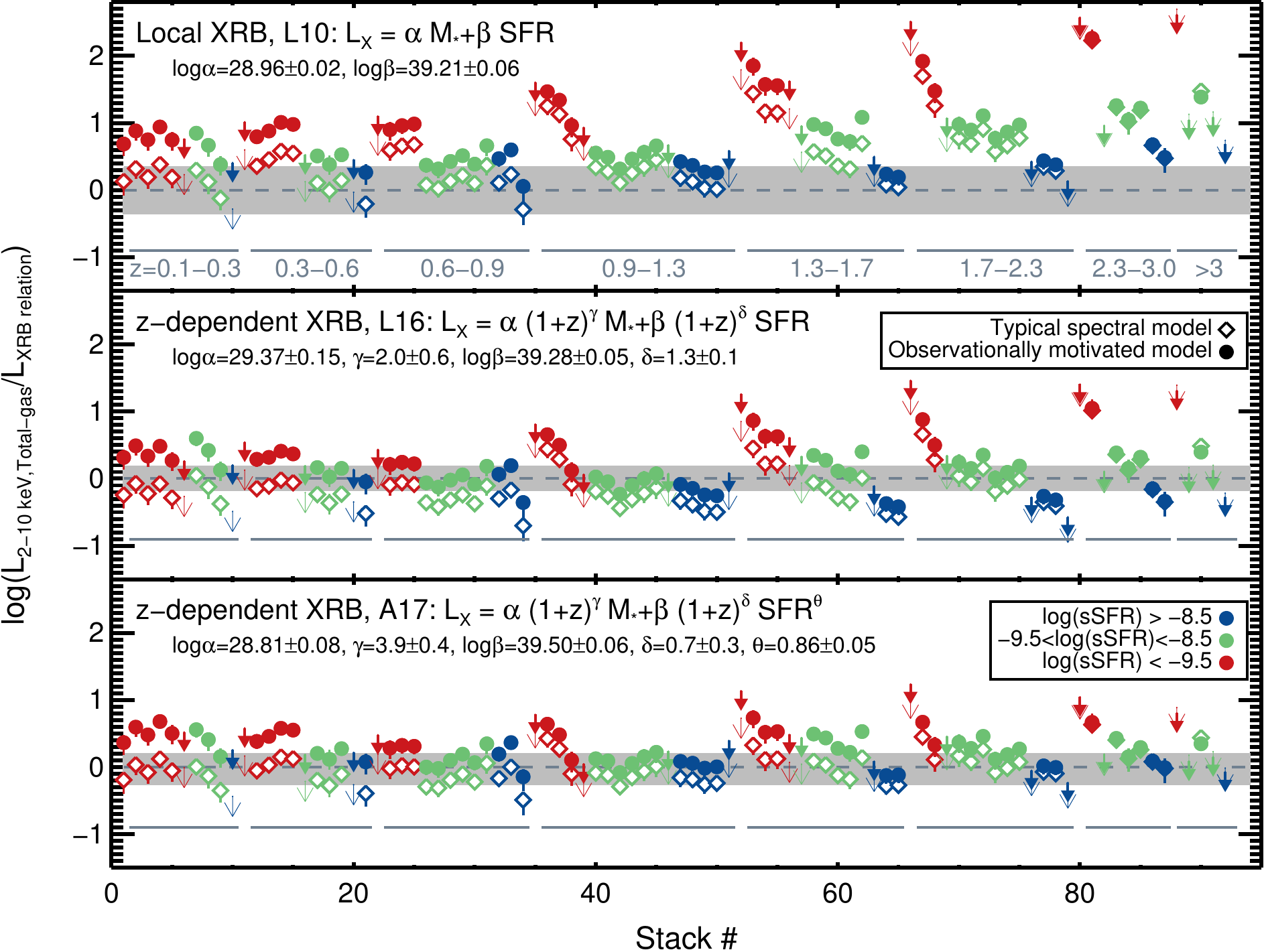}
\caption{The ratio of the rest-frame $2-10$~keV stacked luminosities (from which the hot gas contribution has been subtracted) to the estimated XRB luminosity based on the scaling relations from previous studies are shown for each galaxy stack, placed on the x-axis according to its stack \# from Table \ref{tab:properties}.  Comparisons to the local XRB relation from \citetalias{lehmer10}, the $z$-dependent relation from \citetalias{lehmer16}, and the $z$-dependent relation from \citetalias{aird17a} are shown in the top, middle, and bottom panels, respectively.  Gray areas indicate the observed scatter for the \citetalias{lehmer10} and \citetalias{lehmer16} scaling relations, and the mean statistical error on $L_X^{\mathrm{mode}}$ from \citetalias{aird17a}.  The redshift range of the stacks are shown by labels at the bottom of each panel.  The spectral model used to calculate $\langle L_X\rangle$ and the sSFRs of the galaxy stacks are indicated by the type and color of the symbols as done in Figure \ref{fig:lumdiff}.}
\label{fig:xrb}
\end{figure*}

Having calculated the X-ray luminosity of each stack, we first seek to constrain the contribution of XRBs to the total X-ray emission.  We compared the stacked X-ray luminosities to three XRB emission scaling relations: \par
\noindent (i) the local XRB relation from \citeauthor{lehmer10}(\citeyear{lehmer10}, \citetalias{lehmer10}) computed using a sample of local galaxies and bright infrared galaxies at $z<0.88$ and parametrized so that the LMXB component is proportional to $M_*$ and the HMXB component is proportional to SFR\par
\noindent (ii) the $z$-dependent relation from \citetalias{lehmer16} based on star-forming galaxy stacks up to $z=5$ from the 6~Ms \textit{Chandra} Deep Field South (CDF-S) and the nearby sample from \citetalias{lehmer10}\par
\noindent (iii) the $z$-dependent relation from \citetalias{aird17a} derived from the peaks of X-ray luminosity distributions of star-forming galaxies between $0.1<z<4$ in the COSMOS (\citealt{elvis09}; \citealt{civano16}), AEGIS \citep{nandra15}, CDF-S \citep{xue11}, and CDF-North \citep{alexander03} fields calculated using a Bayesian method.  \par
\noindent All three of these studies adopt spectral models similar to the typical model, and small differences between these models should only result in 0.1 dex variations in the derived X-ray luminosities.  \par 
Figure \ref{fig:xrb} compares the rest-frame $2-10$~keV luminosities of our COSMOS stacks, both those derived using the typical and observationally motivated spectral models, to these three XRB relations.  The hot gas contribution estimated from the SFR-$L_{X,\mathrm{gas}}$ correlation \citep{mineo12b}, which accounts for $<5$\% of the flux in the rest-frame $2-10$~keV band, has been subtracted from the stacked luminosities.  Regardless of the spectral model assumed to derive the X-ray luminosities of our stacks, most low and mid sSFR stacks show excess emission at $z\gtrsim1$ relative to the local \citetalias{lehmer10} relation.  This excess should be considered an upper limit to the nuclear emission in these stacks, because the local XRB relation provides a lower limit to the XRB contribution for galaxies at higher redshift.  The X-ray excess is significantly smaller or even disappears for some stacks if the $z$-dependent XRB scaling relations from \citetalias{lehmer16} or \citetalias{aird17a} are used to estimate the XRB contribution.  As discussed in more detail in \S\ref{sec:hmxbz}-\ref{sec:volume}, the \citetalias{lehmer16} relation overestimates the XRB contribution for most of the high sSFR stacks, especially for those at $z>1$, whereas the \citetalias{aird17a} relation estimates XRB contributions that are either lower than or consistent with the X-ray luminosities of our stacks given the scatter in the relation.   \par
Since the \citetalias{lehmer10} and \citetalias{aird17a} XRB relations assumed unobscured spectral models for the XRB population, they may underestimate the XRB contribution to the stack luminosities if the XRB populations are in fact obscured.  To account for the possibility that the XRB populations are subject to the obscuration levels of our observationally motivated spectral model, we derive a new scaling relation based on our low-redshift stacks, which are the most likely to be XRB-dominated given their low luminosities ($\langle L_X\rangle\lesssim10^{41}$ erg s$^{-1}$) and the fact that their hardness ratios are consistent with the typical spectra of XRBs.  We fit the functional forms of the XRB scaling relations commonly used (\citetalias{lehmer10}; \citetalias{lehmer16}; \citetalias{aird17a}) to the observationally motivated model luminosities of the detected COSMOS stacks at $z<0.9$.  The functional form used by \citetalias{aird17a} ($L_X = \alpha(1+z)^{\gamma}M_*+\beta(1+z)^{\delta}\mathrm{SFR}^{\theta}$) provides the best $\chi^2_{\nu}$ value of 1.3, yielding the following best-fitting parameters:
\begin{equation} 
\label{eq:newxrb}
\begin{split}
log(\alpha) & = 29.98\pm0.12 \\
\gamma & =0.62\pm0.64 \\
log(\beta) & = 39.78\pm0.12 \\
\delta & < 0.2 \\
\theta & = 0.84\pm0.08
\end{split}
\end{equation}
Although this relation is derived using only the COSMOS stacks with $>3\sigma$ significance, it is consistent with the upper limits of stacks at $z<0.9$ with $<3\sigma$ significance. The normalization of this relation is higher and the redshift evolution is weaker than the \citetalias{lehmer16} and \citetalias{aird17a} XRB relations.  \par
The XRB emission estimated using this relation should be considered an upper limit since we cannot completely rule out the presence of low-luminosity AGN in our stacks.  Ideally, in deriving this relation we would also include any individually detected normal (non-AGN) galaxies so as not to introduce a bias towards lower $L_X$ values.  Such a correction is most important in our lowest-redshift bin, where the individual source detection threshold is $L_X\approx1.5\times10^{41}$ erg s$^{-1}$.   We add to our lowest-$z$ stacks all individual detections with $L_X<10^{42}$ erg s$^{-1}$ and an optical/IR counterpart from \citet{marchesi16} which is a star-forming galaxy at $z=0.1-0.3$ \citep{laigle16}.  This inclusion increases the $\langle L_X \rangle$ of the lowest-$z$ stacks by 0.13 dex on average (range of $0.0-0.25$ dex), but the parameters of the XRB relation we derive using these new $\langle L_X \rangle$ values remain statistically consistent.  Therefore, we use the new scaling relation as parametrized in Equation \ref{eq:newxrb} along with the \citetalias{lehmer10} and \citetalias{aird17a} XRB relations to estimate a plausible range for the XRB emission for each stack.  

\section{Discussion}
\label{sec:discussion}
\subsection{The XRB population}
\label{sec:xrbdiscussion}
\subsubsection{Weak redshift evolution of HMXBs}
\label{sec:hmxbz}
The XRB populations in high-sSFR galaxies are dominated by HMXBs, which are correlated with the SFR.  Thus, it is the SFR component of the scaling relations that dominates the estimated XRB contribution for the high-sSFR stacks.  The comparison of the COSMOS high-sSFR stacks with the XRB scaling relations shown in Figure \ref{fig:xrb} reveals that: \par
\noindent(i) most or all of the X-ray emission of the high-sSFR stacks can be attributed to XRBs \par
\noindent(ii) whether or not the X-ray luminosity per SFR of these stacks exhibits substantial redshift evolution depends on the obscuration assumed to calculate the X-ray luminosities.  \par
The $L_X$/SFR of the high-sSFR stacks calculated using our observationally motivated spectral model does not exhibit significant redshift evolution as evidenced by the fact that: \par
\noindent(i) most (12 of 14) high-sSFR stacks are consistent, within the observed scatter, with the local XRB relation \par
\noindent(ii) fitting the high-sSFR stacks with a $z$-dependent model ($L_X=\beta(1+z)^{\delta}$SFR) does not yield a lower $\chi^2_{\nu}$ value than using a $z$-independent model ($L_X=\beta$SFR) \par
\noindent(iii) at $z>1$ these stacks are more consistent with the \citetalias{aird17a} relation than the \citetalias{lehmer16} relation, the former employing a weaker redshift evolution for $L_{\mathrm{XRB}}$/SFR \par
\noindent(iv) our new XRB scaling relation derived by fitting the COSMOS stacks at $z<0.9$ is consistent with no redshift evolution of $L_X$/SFR.  \par
In contrast, looking at the $L_X$/SFR of the high-sSFR stacks calculated using the typical, unobscured spectral model, there is evidence for redshift evolution.  Fitting a $z$-dependent model yields $\chi^2_{\nu}$=1.9 whereas a $z$-independent model results in a much worse $\chi^2_{\nu}$ value of 4.9; in this case, the power-law index of the $z$-dependent model is found to be $1.6\pm0.3$, consistent with that measured by \citetalias{lehmer16}.  The fact that the measurement of HMXB redshift evolution depends on the spectral model used reveals the importance of studying the obscuration of XRBs in star-forming galaxies.  While our observationally motivated spectral model is consistent with the observed hardness ratios of the high-sSFR stacks, and the typical, unobscured model is not, we cannot know whether the observed obscuration is associated with the whole galaxy (and therefore the XRB population) or only the nuclear region (and thus the AGN).  \par
We do not identify any systematic effect that could artificially suppress the redshift evolution of the X-ray luminosities of our high-sSFR stacks.  One systematic effect to consider is the AGN contribution to the stacks and its variation with redshift.  Since the $L_X$ sensitivity threshold above which AGN can be individually detected increases with redshift, higher-$z$ stacks are expected to include more luminous AGN that would instead be excluded from the lower-$z$ stacks.   Thus, we expect the AGN contribution to the stacks to increase with redshift; as a result, the AGN contribution could potentially introduce a positive $z$-dependence to the stacks that might be incorrectly attributed to XRBs, but it is unlikely that it would dilute any positive $z$-evolution that may arise from the XRB population.  Similarly, the possibility that the SED-derived SFRs underestimate the true values (see \S\ref{sec:sample}), which is more likely to occur for higher-$z$ stacks due to their higher SFRs, could produce a false increase of $L_X$/SFR with $z$ but should not produce the opposite trend.  Since HMXBs reside near star-forming regions which are spatially distributed throughout the galaxy rather than being centrally concentrated, another factor to consider is the fractional galaxy area contained within the \textit{Chandra} PSF as a function of redshift.  Given that the PSF extends at least to 2$r_E$ for the vast majority of galaxies, we would not expect a significant fraction of HMXB emission to be missing from our stacks, but even if it were, since the galaxy area covered increases with redshift, the result would be a positive redshift evolution of $L_X$/SFR.  \par
Since all these possible systematics would tend to produce an increase of $L_X$/SFR with $z$, we therefore conclude that the lack of strong redshift evolution of $L_X$/SFR (based on the observationally motivated spectral model) in the high-sSFR stacks is not an artifact.  Thus, assuming that the XRB population is obscured to the extent indicated by our observationally motivated spectral model, these COSMOS stacks suggest that the redshift evolution of HMXB populations is weaker ($\delta<0.2$) than measured by \citetalias{lehmer16} but remains consistent within the scatter of the local HMXB relation \citepalias{lehmer10} and the relation with weak HMXB redshift evolution \citetalias{aird17a}.  Studies of the obscuration of XRB populations in star-forming galaxies as a function of redshift are required to verify this assumption.  
 
\subsubsection{Effect of galaxy sample size on XRB relations}
\label{sec:volume}

The \citetalias{lehmer16} XRB scaling relation significantly overestimates the X-ray luminosity of many high and mid-sSFR COSMOS stacks, measuring a stronger redshift evolution for the $L_X$/SFR of HMXBs than is supported by our data or found by \citetalias{aird17a}.  One substantial difference between our study and the \citetalias{lehmer16} study is that the COSMOS stacks typically contain hundreds to thousands of galaxies, each with relatively shallow exposures of 160~ks, while the CDF-S stacks consist only of tens of galaxies, each with deep exposures of 6~Ms.  Therefore, we investigated whether the differences in galaxy sample size can explain the observed difference between the COSMOS and CDF-S stack luminosities.  \par
For each CDF-S and COSMOS stack, we perform 1000 Monte Carlo simulations.  For each simulation, we randomly select a sample of galaxies from the original COSMOS sample in the $z$, SFR, and $M_*$ range covered by a given stack; the number of galaxies in each random sample is equal to the number of galaxies used in the stack.  Then, we simulate an XRB population for each galaxy based on the HMXB luminosity function (LF) from \citet{mineo12a} and the LMXB LF from \citet{gilfanov04}, which we renormalized to match the local XRB relation from \citetalias{lehmer10}; thus, the number of HMXBs is correlated with the galaxy SFR, the number of LMXBs is correlated with the galaxy $M_*$, and we allow for Poissonian variations in the number of HMXBs and LMXBs.  We calculated the total XRB luminosity for each galaxy by summing the luminosities of HMXBs and LMXBs randomly drawn from their respective LFs.  If the XRB luminosity of a galaxy exceeded the luminosity detection threshold applied to a given stack, we re-simulated the XRB population of the galaxy until its total luminosity fell below the applied threshold; this step was performed in order to maintain the same number of galaxies in our simulated stacks as in our actual stacks.  We calculated the mean XRB luminosity for each stack, and finally, we determined the median and standard deviation of the $\langle L_X\rangle$ distribution of the 1000 simulations of each stack.  \par
We compare the results of these simulations to the expected XRB luminosity of each stack based on the mean SFR and $M_*$ of the stack and the \citetalias{lehmer10} relation.  The simulated luminosities of stacks with $\lesssim$50 galaxies, which includes the vast majority of CDF-S stacks, exceed the expected $\langle L_X\rangle$ based on the \citetalias{lehmer10} relation by 0.15 dex on average.  Such small galaxy samples do not adequately sample the bright end of the HMXB or LMXB luminosity functions, which exhibit steep cutoffs; therefore, a small number of luminous XRBs, or even a single very luminous XRB, can dominate the luminosity of the stack, skewing it to values higher than the $\langle L_X\rangle$ of a well-sampled XRB LF.  \par
The median $\langle L_X\rangle$ of most simulated COSMOS stacks with mid or high sSFR, 63 of which contain $>50$ galaxies, do not show any systematic offset from the expected values.  Thus, the fact that the \citetalias{lehmer16} relation tends to overestimate the XRB contribution of many COSMOS stacks, especially those with mid or high sSFR, can be partly attributed to the differences in galaxy sample size between COSMOS and CDF-S.  \par 
The simulated $\langle L_X\rangle$ of COSMOS stacks with low sSFR are 0.1 dex higher than the expected mean values from the \citetalias{lehmer10} relation, even when they contain as many as 1000 galaxies.  The bright end of the LMXB LF has a steeper power-law index ($-4.8\pm1.1$; \citealt{gilfanov04}) than the HMXB LF ($-2.73^{+0.54}_{-1.58}$; \citealt{mineo12a}); thus, it is not surprising that larger galaxy samples would be required to properly sample the LF of LMXBs, which can dominate the XRB emission of low-sSFR galaxies.  \par
It is important to note that the uncertainties on the bright end slopes of the HMXB and LMXB LFs are large, and therefore improved measurements of the bright end slopes and their evolution with redshift are critical for estimating the galaxy sample sizes necessary to properly measure the redshift evolution of XRB emission.  While a more detailed analysis of this issue based on current theoretical predictions and observational constraints are outside the scope of this paper, our simple simulations suggest that COSMOS-like survey volumes are better suited for studying the correlation of XRB populations with galaxy properties.
\begin{figure*}
\centering
\includegraphics[width=0.85\textwidth]{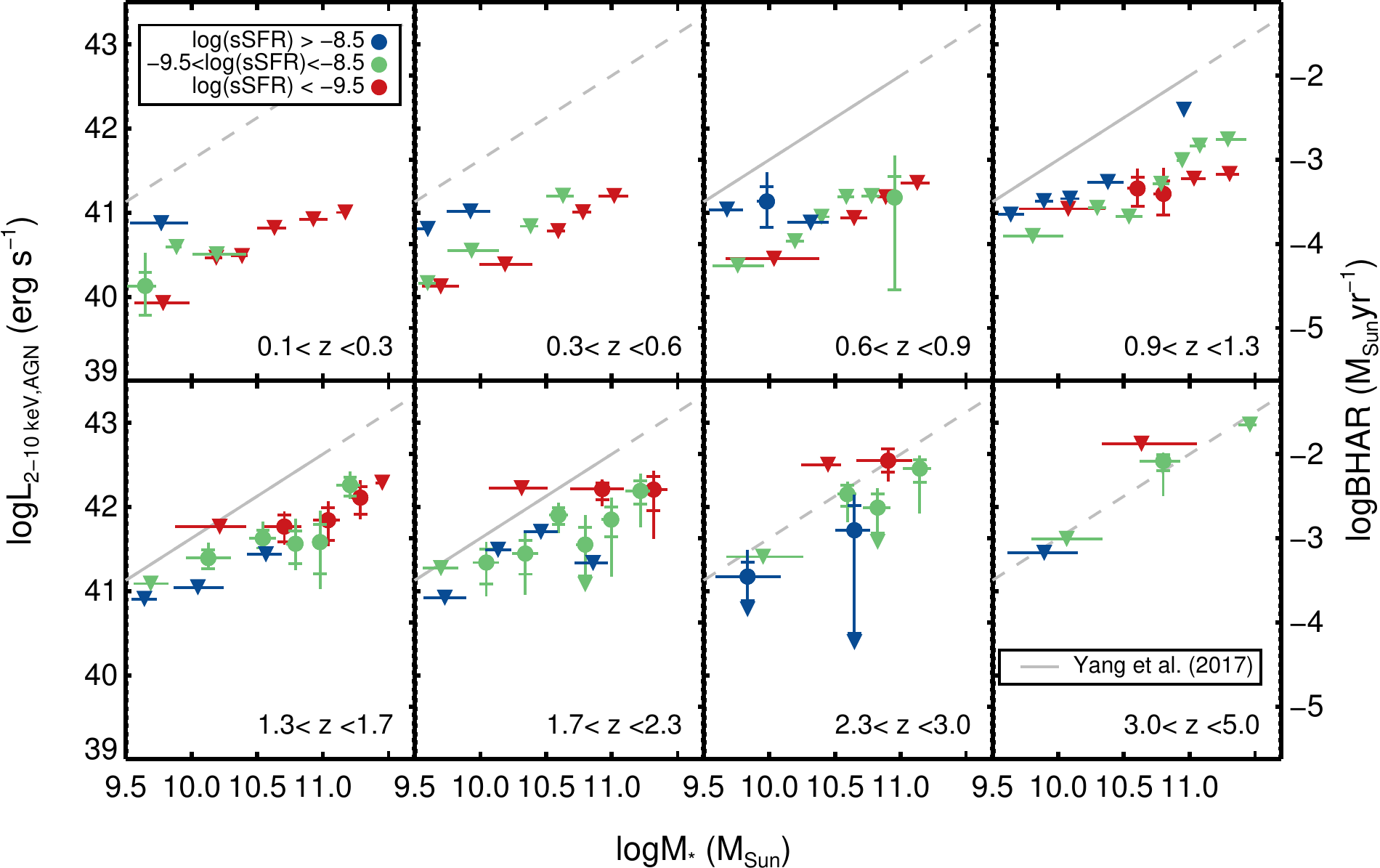}
\caption{The left y-axis shows test-frame $2-10$~keV stacked luminosities attributed to AGN emission (after the gas and XRB contributions have been subtracted) versus the mean stellar mass of each stack as a function of redshift.  The X-ray luminosities shown are based on our observationally motivated spectral model.  The right y-axis shows the black hole accretion rate (BHAR) based on Equation \ref{eq:bhar} versus $M_*$ for each stack as a function of redshift.  The BHAR-$M_*$ relation from \citet{yang17} is shown as a solid line over the $M_*$ and $z$ ranges covered by the galaxy sample used by \citet{yang17} and as a dashed line when it is extrapolated to $M_*$ or $z$ values outside those ranges.  Circles indicate AGN emission calculated by using the new XRB scaling relation parametrized in Equation \ref{eq:newxrb}.  Vertical error bars with hats show $1\sigma$ statistical uncertainty on this estimate of the AGN luminosity.   A maximum limit for the AGN emission calculated by subtracting the hot gas contribution and the XRB contribution based on the \citetalias{lehmer10} XRB relation is shown by a vertical line extending above the statistical error bar.  If no excess AGN emission is estimated when using the new XRB relation, then a triangle symbol indicates the upper limit on the AGN emission.  A minimum limit on the AGN emission calculated by subtracting the hot gas contribution and the XRB contribution based on the \citetalias{aird17a} XRB relation is shown by a vertical line extending below the statistical error bar.  If the minimum limit is equal to zero, this is indicated by a triangle below the statistical error bar.  High, mid, and low sSFR stacks are represented by blue, green, and red symbols, respectively.}
\label{fig:agn}
\end{figure*}

\subsection{The AGN population}
\subsubsection{BH activity in star-forming galaxies}
\label{sec:agnactivity}

In order to study low-luminosity BH activity in star-forming galaxies, we subtract the hot gas and XRB contributions from the luminosities of the COSMOS stacks calculated using the observationally motivated spectral model.  Figure \ref{fig:agn} displays the residual luminosities.  In this figure, the residual $\langle L_X\rangle$ calculated by subtracting the XRB contribution based on the scaling relation parametrized in Equation \ref{eq:newxrb} is shown as a circle with error bars displaying the 1$\sigma$ statistical uncertainty.  We also calculate the residual $\langle L_X\rangle$  and corresponding statistical upper and lower bounds by subtracting the XRB contribution as estimated by the \citetalias{lehmer10} and \citetalias{aird17a} XRB relations; if these upper or lower bounds lie outside the displayed error bars, they are shown by vertical lines extending beyond the error bars.  \par
\begin{figure*}
\centering
\includegraphics[width=0.85\textwidth]{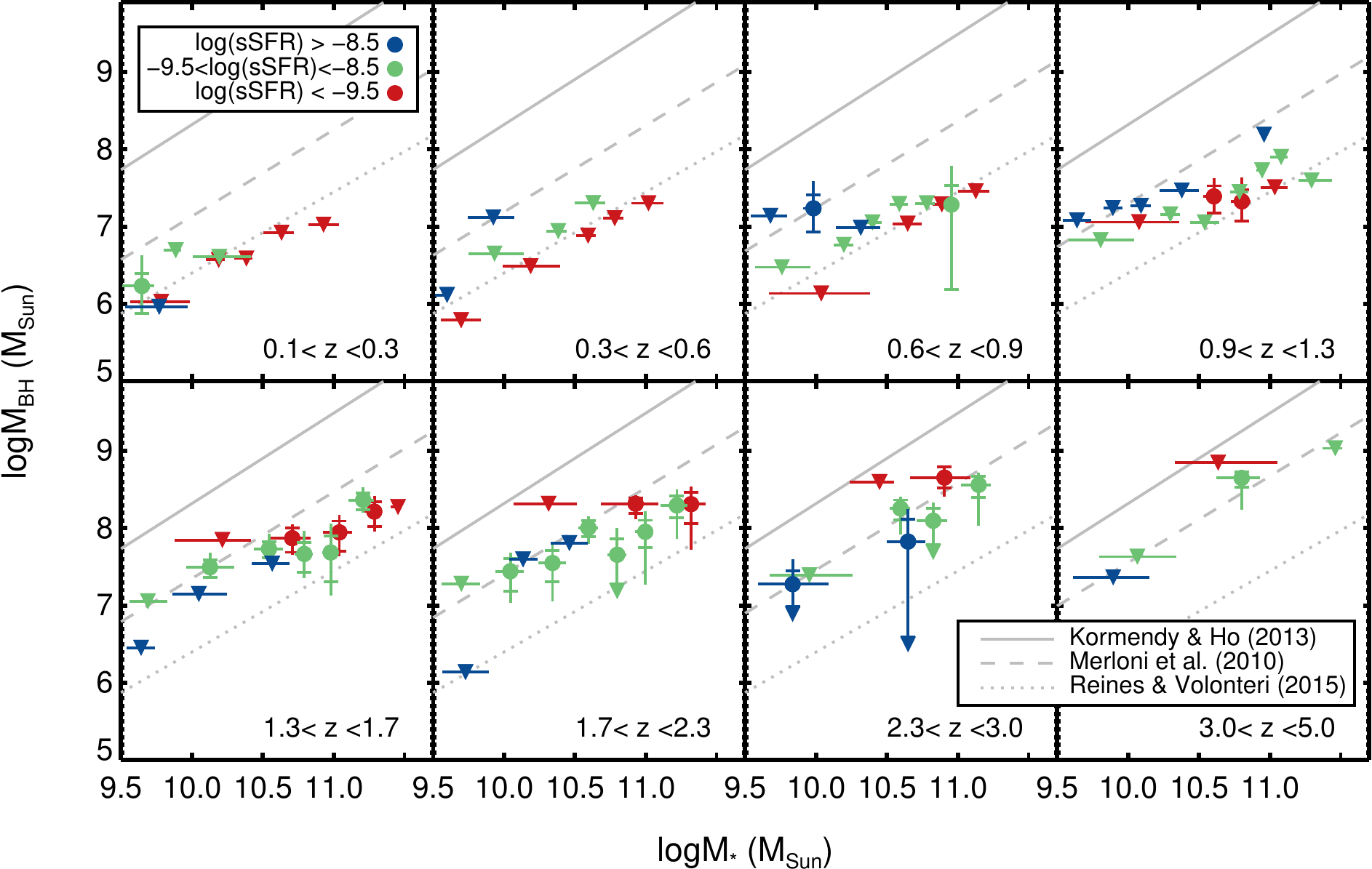}
\caption{Average BH mass versus stellar mass for each stack as a function of redshift.  BH masses are calculated assuming $k_{\mathrm{bol}}=16$ and $\lambda_{\mathrm{Edd}}=10^{-3}$.  Gray lines represent scaling relations from previous studies for comparison.  The symbols and error bars are as described for Figure \ref{fig:agn}.
}
\label{fig:mbh}
\end{figure*}
At $z<1.3$, in all but five stacks, there is no detected residual contribution after accounting for the plausible range of XRB contributions.  Of course, for $z<0.9$, this result is circular logic since we use the $z<0.9$ stacks to derive our new XRB scaling relation which provides one of our XRB contribution estimates.  If we consider only the \citetalias{lehmer10} or \citetalias{aird17a} XRB relations for $z<0.9$, then we find significant residual emission of $\langle L_X \rangle\sim10^{40}-10^{41}$ erg s$^{-1}$ in most $z<0.9$ stacks.  While we are not confident that this emission can be attributed to AGN, we do find that, just like the more robustly detected AGN emission at $z>1.3$ discussed below, this residual $\langle L_X \rangle$ is correlated with $M_*$ but not SFR.  \par
At $z>1.3$ no residual contribution to the high-sSFR stacks is found when accounting for the possible range of XRB contributions, but two thirds of the low-mid sSFR stacks exhibit a significant residual contribution of $10^{41}<L_X<10^{42.5}$ erg s$^{-1}$.  These residual contributions suggest the presence of low-luminosity AGN in these stacks.  As can be seen in Figure \ref{fig:agn}, within each redshift bin above $z=1.3$, the AGN luminosity increases with stellar mass, and when stacks with similar $M_*$ are compared, there does not appear to be a strong correlation between AGN luminosity and SFR.  \par
These results are consistent with recent work by \citet{yang17}, who find that BH activity at $0.5<z<2.0$ is correlated strongly with $M_*$ but only weakly with SFR once the correlation with $M_*$ has been taken into account.  To compare our results more directly with \citet{yang17}, we calculate the mean black hole accretion rate (BHAR) for each stack based on the AGN luminosity and the definition used by \citet{yang17}:
\begin{equation}
\langle \mathrm{BHAR} \rangle = \frac{(1-\epsilon)k_{\mathrm{bol}}\langle L_X \rangle}{\epsilon c^2}
\label{eq:bhar}
\end{equation}
where $k_{\mathrm{bol}}=L_{\mathrm{bol}}/L_X$ is the the bolometric correction factor to convert and $\epsilon$ is the mass-energy conversion efficiency.  Following \citet{yang17}, we assume that $\epsilon=0.1$.  The bolometric correction is luminosity-dependent, but over the AGN luminosity range of the COSMOS stacks ($L_X\sim10^{41}-10^{43}$ erg s$^{-1}$), it only varies from $8-24$ (\citealt{marconi04}; \citealt{hopkins07}; \citealt{lusso12}).  Low-luminosity ($L_X<10^{43}$ erg s$^{-1}$) AGN in local star-forming galaxies have average $k_{\mathrm{bol}}=16$ \citep{she17}, and we opt to use this value for simplicity since the luminosity-dependent corrections would only change the bolometric luminosity by 0.2-0.3 dex.  \par
Figure \ref{fig:agn} shows the mean BHAR of each stack as well as the linear BHAR-$M_*$ relation from \citet{yang17}.  The COSMOS stacks exhibit a roughly linear slope in BHAR-$M_*$ similar to the \citet{yang17} relation, but a different normalization depending on the redshift bin.  The different normalizations are likely due to the fact that: (i) \citet{yang17} combine both X-ray detected AGN and X-ray stacks of undetected galaxies, whereas our stacks do not include X-ray detected AGN and the $L_X$ detection limit varies with redshift, and (ii) \citet{yang17} adopt a higher bolometric correction factor ($k_{\mathrm{bol}}$=22.4) since they include more luminous AGN in their stacks. \par
We test whether the BHAR-$M_*$ relation is correlated with SFR by performing a partial correlation  test using the Kendall $\tau$ coefficient generalized for censored data by \citet{akritas96}.  We perform the test using the detected stacks and upper limits from $z=1.3-2.3$, a redshift range containing most of stacks with residual AGN emission and which is narrow in order to minimize the effect of possible correlations between redshift and other variables (i.e. $M_*$, SFR, and BHAR).  Using the AGN luminosities calculated by subtracting the XRB contribution based on our new relation (see Equation \ref{eq:newxrb}), we find that the BHAR is strongly correlated with $M_*$ ($\tau=0.51$, $p<0.05$) but is only marginally correlated with SFR ($\tau=-0.12$, $p>0.05$).  Performing this test using the local XRB relation or the \citetalias{aird17a} relation to subtract the XRB contribution yields similar results.  \par
The AGN luminosity is proportional to the mass of the supermassive black hole ($M_{\mathrm{BH}}$) and the Eddington ratio ($\lambda_{\mathrm{Edd}}$).  The roughly linear correlation between $M_*$ and $\langle L_X\rangle$ that we measure is consistent with studies which find that the Eddington ratio distribution of AGN is independent of stellar mass for $M_*>10^{10} M_{\odot}$ \citep{aird12,aird17b}.  If the Eddington ratio distribution is independent of $M_*$, then galaxy stacks with different mean $M_*$ should have the same average $\lambda_{\mathrm{Edd}}$.  Then, if $M_{\mathrm{BH}}$ and $M_*$ are linearly correlated as found by several studies (e.g. \citealt{merloni10}; \citealt{reines15}), the average X-ray luminosities and stellar masses of the galaxy stacks should be linearly correlated, as is observed in Figure \ref{fig:agn}. \par   
While we cannot independently constrain $M_{\mathrm{BH}}$ and $\lambda_{\mathrm{Edd}}$, we can check whether reasonable ranges of these parameters, based on previous studies, are consistent with the X-ray luminosities of our COSMOS stacks.  \citet{she17} find that the average $\lambda_{\mathrm{Edd}}$ of AGN in local star-forming galaxies with $L_X<10^{43}$ erg s$^{-1}$ is $\sim10^{-3}$.  Adopting this $\langle\lambda_{\mathrm{Edd}}\rangle$ value, we find that AGN in COSMOS star-forming galaxies have typical BH masses of $10^{7}-10^{9} M_{\odot}$, as shown in Figure \ref{fig:mbh}, and they fall between the \citet{reines15} and \citet{merloni10} $M_{BH}-M_*$ relations, gradually becoming more consistent with the \citet{merloni10} relation as the redshift increases.  The \citet{reines15} relation is based on a sample of nearby AGN spanning a wide range of luminosities ($L_{\mathrm{bol}}=10^{41.5}-10^{44.4}$), while the redshift evolution of the $M_{\mathrm{BH}}-M_*$ relation measured by \citet{merloni10} is based on a sample of luminous ($L_{\mathrm{bol}}=10^{44.5}$ erg s$^{-1}$) type 1 AGN at $1<z<2.2$; thus it seems reasonable for our stacks to approach the \citet{merloni10} relation as the $\langle L_X\rangle$ of the stacks increases with redshift. \par
Assuming the same Eddington ratio ($\langle\lambda_{\mathrm{Edd}}\rangle\sim10^{-3}$), \citet{mezcua16} found that the obscured BHs in COSMOS dwarf galaxy stacks at $z=0.5-1.5$ have lower BH masses of $\sim10^5-10^6 M_{\odot}$, consistent with the lower stellar masses of these galaxies and the \citet{reines15} relation.  If $\langle\lambda_{\mathrm{Edd}}\rangle$ for the COSMOS star-forming galaxies were $10^{-2}$, then the COSMOS stacks would also roughly follow the \citet{reines15} relation for nearby AGN; however, $\langle\lambda_{\mathrm{Edd}}\rangle\sim10^{-2}$ may be an unusually high value since it is estimated that at $1<z<3$ only about 10\% of SMBHs have $\lambda_{\mathrm{Edd}}>10^{-2}$ \citep{aird17b}.  In early-type galaxy stacks at $z\approx1-1.5$, \citet{paggi16} find both highly obscured ($N_{\mathrm{H}}>10^{23}$ cm$^{-2}$) AGN and less obscured ($N_{\mathrm{H}}\sim10^{22}$ cm$^{-2}$) AGN that are consistent with the \citet{kormendy13} relation assuming $\langle\lambda_{\mathrm{Edd}}\rangle\approx10^{-2}$ and $10^{-4}$, respectively.  Our star-forming galaxy stacks would also fall on the \citet{kormendy13} relation if $\langle\lambda_{\mathrm{Edd}}\rangle$ were $10^{-4}$ (under the assumption that $M_{\mathrm{bulge}} = M_*$); this situation would be in sharp contrast to the local Universe, where star-forming galaxies with pseudo-bulges or without bulges fall below the \citet{kormendy13} relation.  \par
The fraction of AGN in a bright state having $\lambda_{\mathrm{Edd}}>10^{-2}$ is referred to as the AGN duty cycle, and it is another factor which can influence how the $\langle L_X \rangle$ of the COSMOS stacks relates to the average AGN BH mass.  If the duty cycle of the AGN in our sample is about 10\%, these bright AGN could dominate the stacked X-ray emission.  In this case, the average $L_X$ of these AGN would be about 10 times higher than the $\langle L_X \rangle$ calculated for all the galaxies in the stack.  In this scenario, we would expect the $M_{\mathrm{BH}}-M_*$ relation for the bright AGN in the COSMOS stacks to be roughly consistent with the \citet{merloni10} relation.  We also note that the duty cycle of the AGN in our sample cannot be significantly less than $10$\%, because in this case, the average $L_X$ of AGN in the bright state would start to exceed the COSMOS sensitivity threshold, and we would expect that such AGN would have been individually detected by \textit{Chandra}.

\subsubsection{Obscured AGN population}
\label{sec:obscured}
As discussed in \S\ref{sec:specmodel}, the hardness ratios suggest that the X-ray sources at $z>1.3$ are highly obscured.   The enhanced obscuration at $z>1.3$ could either be due to higher levels of obscuration throughout the host galaxies or to an increase in the nuclear obscuration fraction.  The hard photon index and X-ray luminosities of the high-redshift stacks with low to mid sSFR indicate that they are dominated by AGN.  \par
The fact that the level of obscuration appears to be higher in the low and mid sSFR stacks than in the high sSFR stacks may be consistent with the trend observed by \citet{lanzuisi17} that the obscured AGN fraction decreases with sSFR and contrary to the positive correlation between obscured fraction and sSFR found by \citet{juneau13}.  However, we cannot exclude the possibility that the high sSFR stacks are dominated by XRBs rather than AGN, and that therefore their lower levels of obscuration are due not to variations in the obscured AGN fraction but the differences between the typical nuclear and galaxy-wide obscuration.  \citet{lanzuisi17} suggest the anti-correlation between obscuration and sSFR may result from a correlation between obscuration and $M_*$ (which is also observed by \citealt{rodighiero15} and \citealt{marchesi16b}) and a lack of correlation between obscuration and SFR.  While the hardness ratios provide a suggestive hint that more massive galaxies are more obscured, it is difficult to untangle this $M_*$ trend from the redshift evolution of the HRs.
\par
\begin{figure*}
\centering
\includegraphics[width=0.85\textwidth]{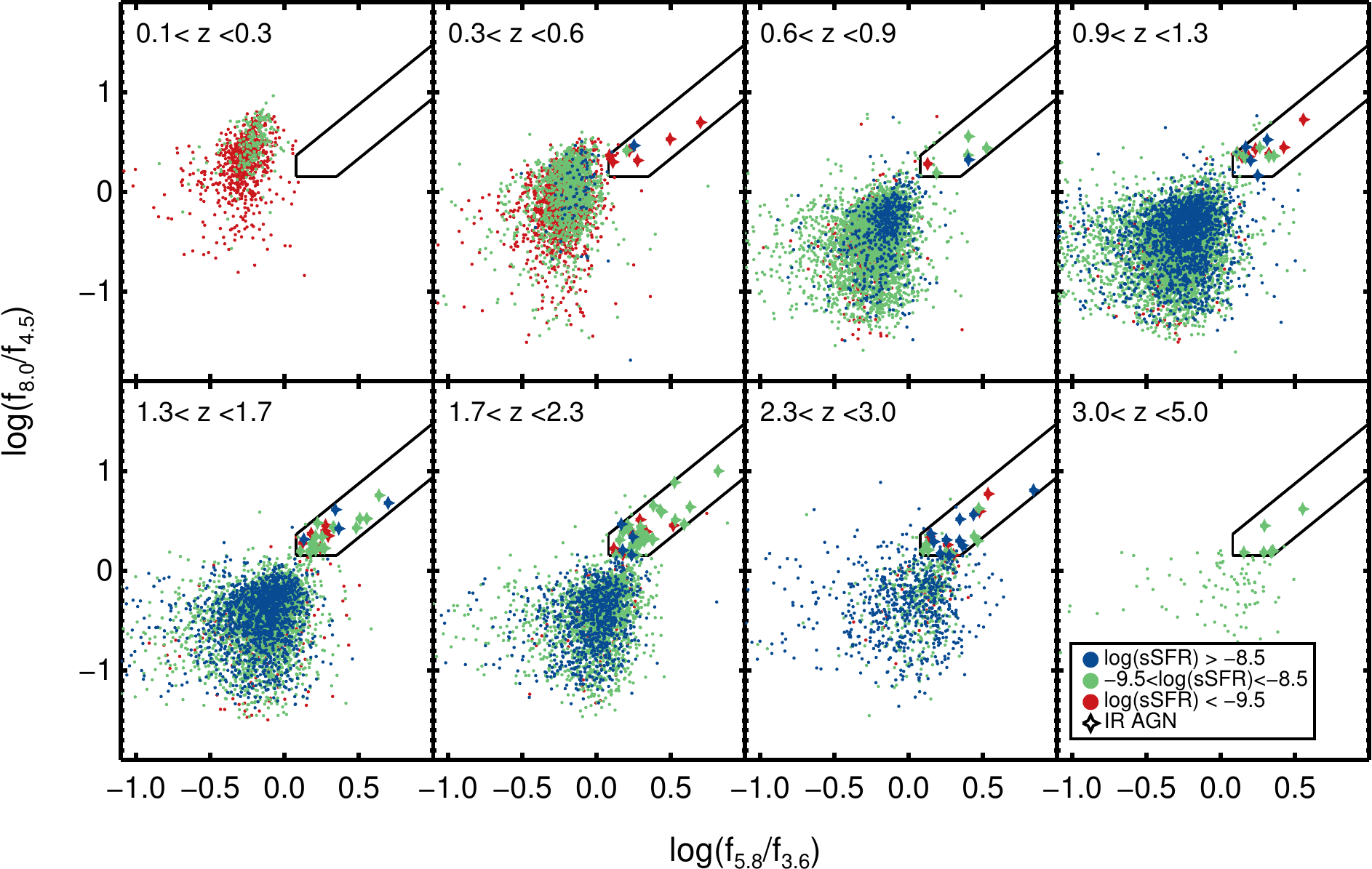}
\caption{IRAC colors from \citet{laigle16} catalog for the galaxies in our significant X-ray stacks as a function of redshift.  Galaxies with high, mid, and low sSFR are represented by blue, green, and red symbols, respectively.  The AGN selection contours from \citet{donley12} are shown with black lines.  Galaxies which meet the IRAC-selection criteria for AGN are shown with diamond symbols.}
\label{fig:iragn}
\end{figure*}
Since evidence for obscured AGN has been found in X-ray stacking studies of star-forming, early-type, and dwarf galaxies in COSMOS at $z\sim1$, it is worth considering what these studies in aggregate reveal about AGN obscuration.  At first glance, these studies seem to indicate that $N_{\mathrm{H}}$ is positively correlated with $M_*$ and anti-correlated with sSFR.  The AGN in dwarf galaxy stacks (log$M_*\approx9-9.5$, sSFR$\sim10^{-8}$ yr$^{-1}$) are obscured by $N_{\mathrm{H}}\sim10^{22}$ cm$^{-2}$ \citep{mezcua16}.  The AGN in star-forming galaxies (log$M_*\approx9.5-11.5$, sSFR$\sim10^{-9}$ yr$^{-1}$) exhibit higher obscuration of $N_{\mathrm{H}}\sim10^{23}$ cm$^{-2}$, and about half of the early-type galaxy stacks (log$M_*\approx10-12$, sSFR$<10^{-11}$ yr$^{-1}$) exhibit even higher obscuration of $N_{\mathrm{H}}\gtrsim10^{23}$ cm$^{-2}$ \citep{paggi16}.  \par
However, this simple picture is complicated by the fact the typical $N_{\mathrm{H}}$ inferred for the AGN in the stacks are towards the upper end of the $N_{\mathrm{H}}$ distribution observed by \citet{lanzuisi17} for individually X-ray detected AGN residing in COSMOS galaxies with similar $M_*$, sSFR, and $z$.  Furthermore, half of the early-type stacks are actually consistent with $N_{\mathrm{H}}\lesssim10^{22}$ cm$^{-2}$ despite having similar $M_*$ and sSFR as the highly obscured early-type stacks \citep{paggi16}.  This suggests that obscuration may depend on an additional parameter besides $M_*$.  \par
\citet{buchner17} propose an AGN obscuration model in which galaxy-wide obscuration depends on $M_*$ and additional nuclear obscuration due to a radiation-lifted torus varies with $\lambda_{\mathrm{Edd}}$.  This model predicts that the nuclear obscuration should peak for $\lambda_{\mathrm{Edd}}\sim10^{-2}$ and decrease at both lower and higher accretion rates.  Such a model could be consistent with the observations of obscured AGN at $z\sim1$. The dwarf, star-forming, and highly obscured early-type galaxy stacks are expected to have $\langle \lambda_{\mathrm{Edd}}\rangle\sim10^{-3}-10^{-2}$ in order to be consistent with observed $M_{\mathrm{BH}}-M_*$ relations.  In contrast, the early-type stacks with low $N_{\mathrm{H}}$ likely have $\lambda_{\mathrm{Edd}}\sim10^{-4}$ \citep{paggi16} while many individually detected AGN may have $\lambda_{\mathrm{Edd}}>10^{-2}$ \citep{lanzuisi17}.  AGN obscuration as a function of host galaxy and SMBH properties will be explored in more detail by Civano et al. (in preparation).

\subsubsection{Comparison with IR AGN selection}
\textit{Spitzer}/IRAC selection has been shown to be effective at identifying obscured AGN of moderate luminosity and can identify AGN not detected by \textit{Chandra} in relatively shallow surveys like COSMOS \citep{donley12}.  Therefore, we test whether the obscured AGN population identified in our X-ray stacks could be discovered through IRAC color selection.  Figure \ref{fig:iragn} displays the IRAC colors of our galaxy sample along with the contours used to identify IR AGN.  As can be seen, only a very small percentage of galaxies (typically $<4$\% of any particular stack) are identified as IR AGN.  Most of the stacks at $z<0.9$ (19 of 28) contain zero IR AGN, consistent with the hypothesis that these stacks are primarily dominated by XRBs.  At $z>1.3$, only three stacks lack IR AGN, while the remaining 24 have IR AGN fractions of 0.1-12\%.  However, for all but 4 of these 27 stacks, if we assume that the IR AGN dominate the stacked X-ray emission, the average $L_X$ of these IR AGN would be comparable to or in excess of the COSMOS sensitivity limit, in which case many of them should have been individually detected by \textit{Chandra}.  \par
We remove the IR-selected AGN from our stacks and recalculate the hardness ratios and X-ray luminosities.  Neither the ratios nor the luminosities are significantly affected by the exclusion of IR AGN, demonstrating that the star-forming stacks are not dominated by a small number of IR AGN.  Thus, there is a population of obscured AGN identifiable in the X-ray stacks at $z>1.3$ that cannot be identified by their IRAC colors.  \par
Evidence for obscured AGN populations have also been found in stacks of dwarf galaxies and early-type galaxies at $z\gtrsim1$ that were not identifiable by their IR colors \citep{mezcua16,paggi16}.  Thus, large near-IR surveys such as those that will be performed by the Wide Field Infrared Survey Telescope (WFIRST) will need to be complemented by X-ray surveys with the next generation of X-ray telescopes to uncover and study the obscured AGN population at high redshift.


\section{Conclusions}
\label{sec:conclusion}
We have performed an X-ray stacking analysis of star-forming galaxies that fall below the \textit{Chandra} COSMOS Legacy survey sensitivity limit in order to study low-luminosity AGN populations, their obscuration, and their connection to host galaxy properties.  Splitting our sample of 75,000 galaxies by redshift, sSFR, and $M_*$ results in 92 bins, 68 of which are detected with $\geq3\sigma$ confidence in the $0.5-2$~keV band.  This X-ray stacking analysis allows us to probe X-ray luminosities as low as $10^{40}-10^{41}$ erg s$^{-1}$ at $z<1.3$ and $10^{41}-10^{42.7}$ erg s$^{-1}$ at $z>1.3$, which are up to two orders of magnitude fainter than the COSMOS sensitivity limit. This study provides insights into both low-luminosity AGN and XRB populations in star-forming galaxies, which are summarized below: \par
1. The hardness ratios and a comparison of the rest-frame $2-10$~keV luminosities derived from the observed $0.5-2$ and $2-8$~keV bands were used to determine an observationally motivated spectral model to convert the stacked count rates into X-ray luminosities.  This spectral model includes substantial obscuration ($N_{\mathrm{H}}\sim10^{22}-10^{23}$ cm$^{-2}$), which increases at $z>1.3$ in low and mid-sSFR galaxies.  The typical spectral models assumed in similar studies of AGN and XRB populations are relatively unobscured ($N_{\mathrm{H}}<10^{21}$ cm$^{-2}$) and result in calculated X-ray luminosities that are up to 0.6 dex lower than those determined using our observationally motivated spectral model.  This difference in the derived luminosities demonstrates the importance of constraining the spectral model in these studies. \par
2. The $L_X$/SFR of the high-sSFR stacks does not exhibit significant redshift dependence.  Since such galaxies are expected to be dominated by HMXBs, our high-sSFR stacks suggest that the redshift evolution of $L_{\mathrm{HMXB}}$ per SFR is weaker than found by \citetalias{lehmer16}, but consistent with the local \citetalias{lehmer10} relation and the weaker redshift dependence found by \citetalias{aird17a}.  This result depends on the obscuration assumed in the spectral model used to derive the X-ray luminosities.  If we adopt an unobscured spectral model, a significant $z$-dependence is measured, but such a spectral model is inconsistent with the observed hardness ratios.   \par
3. We find that, regardless of the spectral model assumed, the X-ray luminosities of our high-sSFR stacks are overestimated by the $z$-dependent XRB scaling relation from \citetalias{lehmer16}. The overestimation of the XRB luminosities of most of the COSMOS high-sSFR stacks and some of the mid-sSFR stacks can be partly explained by the fact that the CDF-S stacks contain too few galaxies to adequately sample the bright end of the XRB luminosity functions, causing them to be biased to higher values.  The COSMOS stacks, which consist of hundreds to thousands of galaxies, are large enough to adequately sample the HMXB luminosity function but are still insufficient to sample the bright end of the LMXB luminosity function, which is very steep. \par
4. The spectral constraints and X-ray luminosities of the low and mid sSFR stacks at $z>1.3$ provide evidence for the presence of an obscured AGN population.  Stacking studies of elliptical and dwarf galaxies in the COSMOS field similarly revealed an obscured AGN population at $z\gtrsim1$.  \par
5. Most of the stacks exhibiting residual AGN emission are at $z=1.3-2.3$.  For the stacks in this redshift range, the average black hole accretion rate increases with $M_*$, but does not show a significant correlation with SFR once the mass dependence has been taken into account.  This result for the low-luminosity AGN population is consistent with results from \citet{yang17}, which includes moderate and high luminosity AGN. \par
6. Assuming the average bolometric correction factor and Eddington ratio for AGN with $L_X<10^{43}$ erg s$^{-1}$ in local star-forming galaxies ($k_{\mathrm{bol}}=16$ and $\lambda_{\mathrm{Edd}}=10^{-3}$), we find that the AGN in our $z>1.3$ stacks have $M_{BH}\sim10^7-10^9 M_{\odot}$.  
The $M_{\mathrm{BH}}-M_*$ relation for these AGN falls between the \citet{reines15} relation for nearby AGN and the $z$-dependent evolution measured for higher-luminosity AGN by \citet{merloni10}.  \par
7. Less than 4\% of the galaxies in each of the COSMOS stacks are identified as AGN via their IRAC colors.  It is not plausible that these IRAC-selected AGN dominate the X-ray emission of the stacks; removing these AGN from our stacks does not significantly impact the measured hardness ratios or X-ray luminosities.  Thus, IRAC color selection is not sufficient for identifying obscured AGN with $L_X<10^{43}$ erg s$^{-1}$.  \par
Future wide, deeper X-ray surveys will be crucial for studying obscured AGN and the evolution of XRB populations.  In order to measure the obscured fraction as a function of AGN luminosity and to reduce systematic uncertainties associated with the spectral model used to calculate X-ray luminosities, a key improvement that could be made by the next generation of X-ray telescopes is to measure spectra, or at least hardness ratios, for individual low-luminosity sources.  \par
The Advanced Telescope for High ENergy Astrophysics (\textit{ATHENA}; \citealt{nandra13}) is an ESA large mission expected to be launched in 2028.  A multi-tier survey strategy being planned for the  \textit{ATHENA} Wide-Field Imager is expected to detect $\sim$400,000 X-ray sources down to $f_X\approx3-7\times10^{-17}$ erg cm$^{-2}$ s$^{-1}$.  Based on the number-count distribution measured in the 7 Ms CDF-S field \citep{luo17}, the vast majority of these sources would be AGN.  These surveys will allow studies of the BH-host galaxy connection and the obscured AGN fraction for $L_X\gtrsim10^{43}$ erg s$^{-1}$ out to $z\sim6$ with AGN samples a factor of $\sim50$ larger than are currently available.  \par
A few thousand normal galaxies are expected to be detected in the \textit{ATHENA} surveys, constituting a factor of $>10$ increase over current samples.  However, since the fluxes of most of these galaxies will be close to the confusion limit of \textit{ATHENA}, their X-ray properties will be difficult to disentangle due to source blending, and it will not be possible to associate many of them to unique multi-wavelength counterparts due to \textit{ATHENA}'s 5$^{\prime\prime}$ PSF.  Attempts to stack \textit{ATHENA} data in order to probe below the sensitivity limit of its surveys will be hampered by its large PSF, which will cause a large fraction of source extraction regions to overlap with one another, biasing the stacked signal.  \par
In comparison, \textit{Lynx}, an X-ray observatory currently being developed in a mission concept study \citep{gaskin17}, would be ideally suited to study both AGN and XRB populations at high redshift, and to connect their properties to their host galaxies.  \textit{Lynx} is planned to have sub-arcsecond resolution over a 20$^{\prime}\times20^{\prime}$ field of view.  This sub-arcsecond resolution would enable the unambiguous association of X-ray sources with their host galaxies down to much lower flux limits than \textit{ATHENA}.  \par
A square degree survey reaching sensitivity limits of $10^{-18}-10^{-19}$ erg s$^{-1}$ cm$^{-2}$ is being considered as part of the \textit{Lynx} concept study.  Based on the CDF-S number-count distribution \citep{luo17}, such a survey would detect about 200,000 sources, $>60$\% of which are likely be normal galaxies, reaching luminosity limits of $L_X\sim10^{41}$ erg s$^{-1}$ out to $z\sim3$ and $L_X\sim10^{42}$ erg s$^{-1}$ out to $z\sim6$.  About half of these sources (including $\approx$60,000 galaxies) would have $>25$ net counts, sufficient for measuring meaningful hardness ratios, and the brightest 25,000 sources (including about $\approx$12,000 galaxies) would have $>100$ net counts, sufficient for simple spectral fitting.  Thus, \textit{Lynx} would be a revolutionary improvement over current studies of low-luminosity AGN and XRBs at high redshift, allowing the individual detection, basic spectral characterization, and unique multi-wavelength association of sources that are currently only X-ray detectable on average through stacking analysis.
\acknowledgments

We thank the referee for comments which improved the clarity and robustness of the results.  We thank M. Mezcua for fruitful conversations, especially regarding AGN obscuration.  The scientific results reported in this article are based on observations made by the \textit{Chandra X-ray Observatory}.  F.F. acknowledges support from NASA grant 44A-1097631.  T. M. and the development of CSTACK are supported by UNAM-DGAPA PAPIIT IN104216 and CONACyT 252531.  A.Z. acknowledges funding from the European Research Council under the European Union?s Seventh Framework Programme (FP/ 2007?2013)/ERC Grant Agreement no. 617001.  This project has received funding from the European Union?s Horizon 2020 research and innovation program under the Marie Sklodowska-Curie RISE action, grant agreement no. 691164 (ASTROSTAT).

\newpage
\clearpage

\centering
\begin{longtable}{cccccccccp{0.5in}}
\kill
\caption{Properties of Star-forming Galaxy Stacks} \\
\hline \hline
\T Stack \# & \# Galaxies & $\langle z \rangle$ & $z$ range & $\langle$log($\frac{M_*}{M_{\odot}}$)$\rangle$ & log($\frac{M_*}{M_{\odot}}$) range & $\langle$log($\frac{\mathrm{SFR}}{M_{\odot} \mathrm{yr}^{-1}}$)$\rangle$ & log($\frac{\mathrm{SFR}}{M_{\odot} \mathrm{yr}^{-1}}$) range & $\langle$log($\frac{\mathrm{sSFR}}{\mathrm{yr}^{-1}}$)$\rangle$\\
\B(1) & (2) & (3) & (4) & (5) & (6) & (7) & (8) & (9) \\

\hline
\endfirsthead
\caption{Properties of Star-forming Galaxy Stacks (continued)} \\
\hline
\T \B Stack \# & \# Galaxies & $\langle z \rangle$ & $z$ range & $\langle$log($\frac{M_*}{M_{\odot}}$)$\rangle$ & log($\frac{M_*}{M_{\odot}}$) range & $\langle$log($\frac{\mathrm{SFR}}{M_{\odot} \mathrm{yr}^{-1}}$)$\rangle$ & log($\frac{\mathrm{SFR}}{M_{\odot} \mathrm{yr}^{-1}}$) range & $\langle$log($\frac{\mathrm{sSFR}}{\mathrm{yr}^{-1}}$)$\rangle$\\
\hline
\endhead
\\
\endfoot
\endlastfoot
1&483&0.22&$0.1-0.3$&9.78$^{+0.20}_{-0.22}$&$9.50-10.10$&-0.20$^{+0.37}_{-0.44}$&$-1.30-0.55$&-9.98$^{+0.31}_{-0.45}$\\
2&111&0.23&$0.1-0.3$&10.19$^{+0.04}_{-0.09}$&$10.10-10.30$&0.13$^{+0.44}_{-0.55}$&$-0.80-0.72$&-10.06$^{+0.39}_{-0.47}$\\
3&74&0.23&$0.1-0.3$&10.38$^{+0.04}_{-0.08}$&$10.30-10.50$&0.23$^{+0.36}_{-0.36}$&$-0.60-0.93$&-10.16$^{+0.33}_{-0.36}$\\
4&68&0.24&$0.1-0.3$&10.63$^{+0.09}_{-0.13}$&$10.50-10.80$&0.30$^{+0.42}_{-0.32}$&$-0.43-0.93$&-10.33$^{+0.40}_{-0.39}$\\
5&18&0.24&$0.1-0.3$&10.93$^{+0.11}_{-0.11}$&$10.81-11.10$&0.52$^{+0.25}_{-0.15}$&$0.00-0.90$&-10.41$^{+0.18}_{-0.12}$\\
6&2&0.24&$0.1-0.3$&11.17$^{+0.03}_{-0.07}$&$11.10-11.20$&0.67$^{+0.03}_{-0.07}$&$0.60-0.70$&-10.50$^{+0.00}_{-0.00}$\\
7&108&0.24&$0.1-0.3$&9.65$^{+0.08}_{-0.14}$&$9.50-9.80$&0.41$^{+0.22}_{-0.23}$&$0.01-0.94$&-9.24$^{+0.19}_{-0.22}$\\
8&46&0.24&$0.1-0.3$&9.89$^{+0.04}_{-0.09}$&$9.80-9.98$&0.65$^{+0.12}_{-0.21}$&$0.40-1.25$&-9.24$^{+0.14}_{-0.23}$\\
9&41&0.23&$0.1-0.3$&10.19$^{+0.22}_{-0.19}$&$10.00-10.59$&0.90$^{+0.20}_{-0.24}$&$0.60-1.50$&-9.29$^{+0.16}_{-0.19}$\\
10&4&0.25&$0.1-0.3$&9.77$^{+0.20}_{-0.24}$&$9.53-9.97$&1.46$^{+0.28}_{-0.22}$&$1.24-1.75$&-8.31$^{+0.18}_{-0.10}$\\
\hline \hline
\multicolumn{9}{p{7.0in}}{\T Notes: Sample of Table 2.  Full table available online.  Errors represent $1\sigma$ standard deviations of galaxy properties for the stacked galaxy samples.}
\label{tab:properties}
\end{longtable}

\centering
\begin{longtable}{ccccccccp{0.5in}}
\kill
\caption{Stacked Photometry} \\
\hline \hline
\T Stack \# & Sig. & Exp. (ks) & Net Counts & $\langle$log($\frac{L_{X,\mathrm{obs}}}{\mathrm{erg s}^{-1}}$)$\rangle$ & $\langle$log($\frac{L_{X,\mathrm{typ}}}{\mathrm{erg s}^{-1}}$)$\rangle$ & log($\frac{L_{\mathrm{AGN}}}{\mathrm{erg s}^{-1}}$) & log($\frac{L_{\mathrm{AGN}}}{\mathrm{erg s}^{-1}}$)  range & log($\frac{L_{\mathrm{XRB}}}{\mathrm{erg s}^{-1}}$)\\
\B(1) & (2) & (3) & (4) & (5) & (6) & (7) & (8) & (9) \\
\hline
\endfirsthead
\caption{Stacked Photometry (continued)} \\
\hline
\T \B Stack \# & Sig. & Exp. (ks) & Net Counts & $\langle$log($\frac{L_{X,\mathrm{obs}}}{\mathrm{erg s}^{-1}}$)$\rangle$ & $\langle$log($\frac{L_{X,\mathrm{typ}}}{\mathrm{erg s}^{-1}}$)$\rangle$ & log($\frac{L_{\mathrm{AGN}}}{\mathrm{erg s}^{-1}}$) & log($\frac{L_{\mathrm{AGN}}}{\mathrm{erg s}^{-1}}$)  range & log($\frac{L_{\mathrm{XRB}}}{\mathrm{erg s}^{-1}}$)\\
\hline
\endhead
\\
\endfoot
\endlastfoot
1&3.4&37655.0&59.9$^{+17.1}_{-17.1}$&39.89$^{+0.11}_{-0.18}$&39.34$^{+0.10}_{-0.20}$&$-$&$<$39.93&40.03;39.20;39.52\\
2&5.3&8265.1&43.8$^{+10.3}_{-9.2}$&40.44$^{+0.08}_{-0.15}$&39.88$^{+0.08}_{-0.15}$&$<$39.92&$<$40.47&40.39;39.55;39.84\\
3&4.4&5242.7&28.9$^{+8.4}_{-7.4}$&40.45$^{+0.11}_{-0.16}$&39.89$^{+0.10}_{-0.17}$&$-$&$<$40.48&40.55;39.69;39.96\\
4&7.5&5053.1&56.4$^{+10.2}_{-9.1}$&40.79$^{+0.07}_{-0.08}$&40.24$^{+0.06}_{-0.08}$&$<$40.20&$<$40.82&40.76;39.85;40.11\\
5&4.8&1658.9&21.0$^{+6.5}_{-5.4}$&40.86$^{+0.12}_{-0.16}$&40.31$^{+0.12}_{-0.17}$&$-$&$<$40.92&41.03;40.11;40.36\\
6&0.9&140.1&1.3$^{+2.7}_{-1.3}$&$<$41.09&$<$40.56&$-$&$<$41.00&41.26;40.32;40.56\\
7&5.3&9036.2&52.6$^{+10.8}_{-10.8}$&40.51$^{+0.08}_{-0.11}$&39.97$^{+0.07}_{-0.12}$&40.15$^{+0.16}_{-0.33}$&$39.82-40.53$&40.26;39.66;39.95\\
8&4.1&3825.7&27.0$^{+8.3}_{-7.3}$&40.57$^{+0.11}_{-0.17}$&40.04$^{+0.10}_{-0.18}$&$<$40.24&$<$40.59&40.47;39.90;40.15\\
9&3.9&2725.8&20.3$^{+7.0}_{-5.9}$&40.55$^{+0.13}_{-0.16}$&40.06$^{+0.13}_{-0.16}$&$-$&$<$40.50&40.71;40.16;40.38\\
10&1.8&349.0&4.0$^{+3.6}_{-2.4}$&$<$41.10&$<$40.44&$<$40.15&$<$40.87&41.03;40.68;40.83\\
\hline \hline
\multicolumn{9}{p{7.0in}}{\T Notes: Sample of Table 3.  Full table available online.  All errors represent $1\sigma$ uncertainty.
(2) Significance of the stack in the $0.5-2$~keV band.
(4) Background-subtracted net counts.  
(5) Mean X-ray luminosity based on the observationally motivated spectral model which varies with $z$ and sSFR.  For stacks with $<3\sigma$ significance, the 3$\sigma$ upper limit is provided.
(6) Mean X-ray luminosity based on the typical spectral model with $\Gamma=1.9$ and only Galactic absorption.  For stacks with $<3\sigma$ significance, the 3$\sigma$ upper limit is provided.
(7) AGN luminosity calculated by subtracting the hot gas contribution and XRB contribution based on our XRB scaling relation parametrized in Equation \ref{eq:newxrb} from $\langle L_{X,\mathrm{phys}} \rangle$ in column 5.  For stacks with $>3\sigma$ significance, if the XRB and gas contributions exceed $\langle L_{X,\mathrm{phys}} \rangle$, 1$\sigma$ upper limit is shown; if the XRB and gas contributions exceed even the $1\sigma$ upper limit on $\langle L_{X,\mathrm{phys}} \rangle$ (or the $3\sigma$ upper limit for stacks with $<3\sigma$ significance), the column is blank.  
(8) Minimum and maximum limit on AGN emission based on statistical errors and accounting for plausible range of XRB contribution based on three values in column 9.
(9) Estimated XRB contribution based on our XRB scaling relation parametrized in Equation \ref{eq:newxrb}, the \citet{lehmer10} relation, and the \citet{aird17a} relation.}
\label{tab:photometry}
\end{longtable}

\end{document}